\documentclass[10pt]{article}

\usepackage{amsmath}
\usepackage{amssymb}

\usepackage{graphicx}

\usepackage{cite}

\usepackage{color} 
\usepackage{amsmath,amssymb,graphics,graphicx} 

\usepackage[pdftex,bookmarks,colorlinks,breaklinks]{hyperref}  
\usepackage{setspace}

\usepackage[labelfont=bf,labelsep=period,justification=raggedright]{caption}

\makeatletter
\renewcommand{\@biblabel}[1]{\quad#1.}
\makeatother

\date{}

\begin{document}

\begin{flushleft}
{\Large
\textbf{Digit patterning during limb development as a result of the BMP-receptor interaction}
}
\\[0.25cm]
Amarendra Badugu$^{1}$, 
Conradin Kraemer$^{1}$, 
Philipp Germann$^{1,2}$, 
Denis Menshykau$^{1}$, 
Dagmar Iber$^{1,2,\ast}$
\\
\bf{1} Department for Biosystems Science and Engineering (D-BSSE), ETH Zurich, Basel, Switzerland
\\
\bf{2} Swiss Institute of Bioinformatics \\[1cm]
\end{flushleft}

\noindent Correspondence and requests for materials should be addressed to D.I. (dagmar.iber@bsse.ethz.ch) \\[1cm]


\section*{Abstract}
Turing models have been proposed to explain the emergence of digits during limb development. However, so far the molecular components that would give rise to Turing patterns are elusive. We have recently shown that a particular type of receptor-ligand interaction can give rise to Schnakenberg-type Turing patterns, which reproduce patterning during lung and kidney branching morphogenesis. Recent knock-out experiments have identified Smad4 as a key protein in digit patterning. We show here that the BMP-receptor interaction meets the conditions for a Schnakenberg-type Turing pattern, and that the resulting model reproduces available wildtype and mutant data on the expression patterns of BMP, its receptor, and Fgfs in the apical ectodermal ridge (AER) when solved on a realistic 2D domain that we extracted from limb bud images of E11.5 mouse embryos. We propose that receptor-ligand-based mechanisms serve as a molecular basis for the emergence of Turing patterns in many developing tissues.

\newpage

\section*{Introduction}
The limb has long served as a paradigm to study organ development. Its easy access enabled developmental biologists to use tissue grafting to define organizing centers, and advanced genetics has led to the identification of the key regulatory circuits \cite{Benazet:2009p41334,Zeller:2009p24283}. Early grafting experiments showed that tissue from the posterior limb bud (referred to as zone of polarizing activity, ZPA) induced ectopic digits when implanted in the anterior limb bud \cite{Tickle:1981p21603}. These extra digits formed a mirror image of the normal forelimb, i.e. 1234554321 in mice and 234432 in chicken. Here numbers refer to digits with the lowest number referring to the most anterior digit in the wildtype (i.e. the thumb in humans). The effect was shown to be concentration-dependent in that a smaller number of grafted cells induced fewer digits and failed to induce the most posterior identities in the anterior limb bud \cite{Tickle:1981p21603}. Wolpert's French Flag model explained these experiments by suggesting that digit number and identity were determined by the local concentration of a chemical compound \cite{Wolpert:1969p21589}. This compound was predicted to be produced in the ZPA and to diffuse across the limb bud, thereby creating a concentration gradient. Decades after the grafting experiments had suggested that patterning was induced by a diffusible compound, Sonic hedgehog (SHH) was isolated and shown to affect patterning in a concentration-dependent manner \cite{Riddle:1993p21612, LopezMartinez:1995p21648}. In agreement with the morphogen model, SHH is expressed in cells of the ZPA and diffuses across the limb bud \cite{Harfe:2004p18349}. In mice that do not express SHH only digit 1  develops \cite{Kraus:2001p18579}, and in the polydactylous mouse mutant extra toes (Xt) SHH is expressed also in the anterior region of the limb bud \cite{Buscher:1997p13518}. 

However, several other experiments question this simple morphogen model and suggest that a more complex mechanism determines digit patterning during limb bud development. For one, polydactyly is still observed even when SHH expression is abolished in the extra toe (Xt) mutant \cite{Litingtung:2002p13571}. Moreover, an experiment in which SHH expression is stopped prematurely at different stages during limb development results in an alternating anterior-posterior sequence of digit formation that cannot be reconciled with the standard morphogen model \cite{Zhu:2008p14377}. Thus the shortest pulse that yielded an additional digit induced formation of digit 4, followed by digit 2, and then digits 3 and 5. Based on the morphogen model digits would be expected to emerge simultaneously or in an anterior-posterior sequence.

Experiments in which SHH-expressing cells were marked revealed that descendants of the SHH-producing cells could be found far beyond the small posterior zone in which SHH expression can be detected \cite{Harfe:2004p18349, Ahn:2004p14441}. Descendants of the SHH-producing cells filled almost half of the limb bud and encompassed all cells in the two most posterior digits and also contributed to the middle digit \cite{Harfe:2004p18349, Ahn:2004p14441}. Based on this observation it was suggested that it is the length for which cells express and secrete SHH (and therefore experience a high SHH concentration) that determines digit specification \cite{Harfe:2004p18349, Ahn:2004p14441}. This model, however, still cannot explain the observed sequence of digit formation.

A number of mathematical models have been developed to explain limb bud development and digit patterning. Based on experimental observations, the earliest model defined some 40 years ago a set  of simple rules to recapitulate the observed changes in the limb buds' size and shape \cite{Ede:1969ua}. The modelling effort highlighted the importance of oriented cell division to obtain the  elongated shape that is characteristic  of the limb bud. Recent 3D imaging data in combination with a Navier-stokes description of the tissue as an viscous fluid indeed confirmed that isotropic proliferation alone is insufficient to explain the shape and size changes during limb bud development \cite{Boehm:2010p42421}. Local gain rates much higher than the measured proliferation rates as well as loss rates had to be included; these may reflect oriented cell division and directed cell migration. The description of the limb tissue as a viscous fluid was first introduced by Dillon and Othmer who studied a model where the local growth rate (the source term in the Navier stokes equation) depended on the concentration of two signaling factors, SHH and FGF \cite{Dillon:1999p12911}. A more recent model for limb bud growth explores the impact of differential tissue elasticity \cite{Morishita:2008p28252}. Other models have focused on particular regulatory interactions that pattern the limb expression domains, i.e. the AER-ZPA regulatory interaction \cite{Hirashima:2008p12949}, the establishment of the proximal and distal signaling centers \cite{Probst:2011jo},  and the diffusivity of SHH \cite{Dillon:2003p12910}.  

Most models, however, have focused on the skeletal patterns that emerge during limb bud development \cite{Limbpatternphysic:2009fw,Alber:2008p28312,Miura:2006bd,Hentschel:2004p39051,Miura:2004jz,Oster:1985p39183,Oster:1983wf, Izaguirre:2004hl, Chaturvedi:2005jj, Christley:2007p12953}. A number of different model types have been studied ranging from continuous to cell-based models and from rule-based stochastic models to reaction-diffusion models; for reviews see \cite{Maini:1991us, Newman:2008p12951}. Turing mechanisms are a particular type of reaction-diffusion models \cite{Turing:1952p868} and have been  frequently  invoked to explain the emergence of pattern during development \cite{Kondo:2010bx}; they have been studied in the context of limb patterning since 1979 \cite{Newman:1979vo}. In Turing models the number of modes (patterns) increases with larger domain size. Newman and co-workers showed that the different number of skeletal elements in the stylopod, zeugopod,  and autopod in anterior-posterior direction can in principle be achieved as a result of the increase in the proximal-distal length; here they have to assume that patterns emerge sequential in stylopod, zeugopod, and autopod and are fixed (frozen) upon their emergence \cite{Hentschel:2004p39051}. Newman and co-workers further showed that parameter values can be identified that give realistic steady state pattern if the equations are solved on static domains whose shapes are based on the limb shapes of various fossils \cite{Zhu:2010km}. However, there is no evidence that the patterning domain in the embryo has the same shape as the adult fossil and patterning is well  known to occur on a growing rather than static domain. Miura and co-workers explored how the speed of pattern emergence could be affected \cite{Miura:2004jz} and studied Turing patterning on growing limb domains \cite{Miura:2006bd}. In particular, Miura and co-workers show that a Turing model can explain the increased number of digits in the doublefoot mutant with the observed increased domain size \cite{Miura:2006bd}.

The many computational studies show that Turing models, in principle, are sufficiently flexible to account for the experimentally observed pattern in wildtype and mutant limb buds. However, in spite of the demonstrated sensitivity of Turing models to domain size and particular reaction types and kinetics the models have so far not been solved on realistic domains with biochemically realistic reaction kinetics. Recent efforts in this direction have been made by the Newman group who propose TGF-beta family ligands as the activator and FGFs, Noggin, Notch and/or CHL2 as the inhibitor in their Turing models with Schnakenberg kinetics \cite{Hentschel:2004p39051, Alber:2008p28312,Zhu:2010km}. However, mutant mice with defects in TGF-beta \cite{Spagnoli:2007eg}, FGF \cite{Lewandoski:2000p14621,Moon:2000p14615,Sun:2000p22667}, Noggin \cite{Brunet:1998p28177}, and Notch \cite{Xu:2010kv} signaling all have digits, while CHL2 expression has been found to be restricted to chondrocytes of various developing joint cartilage surfaces and connective tissues in reproductive organs \cite{Nakayama:2004kp}. While it is possible that the apparent robustness of the patterning mechanism to these mutations is a consequence of redundant regulatory interactions, we wondered whether the reported regulatory interactions would allow us to construct a model that would be consistent with available experimental data and that would not have to rest on missing data and redundancy arguments. 

Advances in experimental techniques now provide us with detailed information regarding the domain geometries \cite{Boehm:2010p42421}, the timing of the processes, and the regulatory interactions  \cite{Towers:2009p18309, Benazet:2009p41334,Zeller:2009p24283}. Limb buds first form around embryonic day (E)9.5 while expression of Sox9, an early marker for tissue condensations, is first observed around E10.5 in the central mesenchyme of the limb bud (Fig. \ref{fig2}A,  \cite{Kawakami:2005p44294}). Around E11 the Sox9 expression pattern splits and a region in the stalk patterns the stylopod, two patches mark the zeugopod, and a region of strong expression marks the autopod (Fig. \ref{fig2}B,F, \cite{Kawakami:2005p44294,Stricker:2011cm}). Around E11.5 the autopod expression pattern splits into distinct patches that mark the different digits (Fig. \ref{fig2}C,G, \cite{Kawakami:2005p44294,Stricker:2011cm}). Over the next day more defined and elongated Sox9 expression pattern emerge that resemble the final digit structure (Fig. \ref{fig2}D,E,H \cite{Kawakami:2005p44294,Stricker:2011cm}). Sox9 expression has been shown to be stimulated by BMP-2  \cite{Healy:1999p44253} and the BMP-2 dependent increase in Sox9 then triggers increased expression of Noggin  \cite{Zehentner:2002p44310} which acts as an antagonist of BMP signaling by sequestering BMP in inactive complexes \cite{Zimmerman:1996vv}. Noggin null mice form digits, but the digit condensations are much wider than in wildtype \cite{Brunet:1998p28177}. Much as Sox9, Noggin is expressed in digit condensations and serves as a marker of endochondral differentiation \cite{Brunet:1998p28177}. 

Even though BMPs also induce the expression of the BMP antagonist Gremlin \cite{Pereira:2000vn}, Gremlin (unlike Noggin) is expressed in the interdigital space rather than in the digit condensations \cite{Weatherbee:2006p41231}.  Another study notes a concentration-dependent regulation of Gremlin expression in that low concentrations of BMP2 upregulate Gremlin while high concentrations of BMP2 downregulate Gremlin in limb mesenchyme cultures \cite{Nissim:2006p27803}. It is thus possible that stronger BMP signaling in the digit condensations induces Sox9 while weaker BMP signaling in the interdigital space induces Gremlin. Similarly, it has been suggested that low BMP concentrations support expression of FGF8 while high BMP concentrations repress 
FGF4 and FGF8 expression in the apical ectodermal ridge (AER) \cite{Liu:2005fr, PajniUnderwood:2007p13328}. 
Two different BMP receptors are expressed in the limb mesenchyme, BMPR1a (ALK3) and BMPR1b (ALK6) that bind the different BMP ligands with different affinities. BMPR1a binds mainly BMP4 while BMPR1b binds BMP4, BMP2 and less well to BMP7 \cite{tenDijke:1994vv,Yamaji:1994bn}. The BMP4-BMPR1b affinity has been established in a Scatchard plot as Kd = 530 pM \cite{Yamaji:1994bn}. Activin receptors (ActRI/ALK2 and ActRII) are also expressed in the limb mesenchyme \cite{Verschueren:1995tm} but appear to bind their BMP ligands with lower affinity \cite{Greenwald:2003tj}. BMP7 binds ActRII with 3-fold higher affinity compared to BMP2 (Kd = 1.7 $\mu$M versus  5.4 $\mu$M) and binds ActRI/ALK2 only with low affinity  (Kd = 143 $\mu$M) \cite{Greenwald:2003tj}. It should be noted that the values are based on Biacore measurements and thus do not adequately reflect the cooperative binding effects.

In both mouse and chicken expression of BMPR1b is restricted to precartilaginous condensations \cite{Baur:2000wb, Montero:2008bp} while Bmpr1a is expressed at low levels throughout the limb bud mesenchyme \cite{Yoon:2005ka, Montero:2008bp}. In the chicken Bmpr1a expression is elevated at the border between precartilaginous condensations and mesenchyme, but low or absent in the precartilaginous condensations. Accordingly,  overexpression of dominant negative (DN)-BRK-2 (the chicken receptor that corresponds to the mouse BMPR1B receptor) blocks chondrogenesis, but not overexpression of (DN)-BRK-1 (which corresponds to mouse BMPR1A) or DN-ActR, 1  \cite{Kawakami:1996vo}. BMPR1B thus appears to be the major transducer of BMP signals in chicken limb condensations. 

In the chicken the two receptors have been shown to be engaged in different regulatory interactions. Thus BMP beads induce noggin, tgfb2, and bmpr1b expression only when implanted into the tip of the chicken digit, but not when implanted in the interdigital space. Also bmpr1a is not expressed when BMP beads are implanted in the interdigital space while expression of the bmpR-1b gene was enhanced in the adjacent digits with the spatial distribution appearing displaced toward the margin of the digits adjacent to the bead \cite{Merino:1998ha}. It therefore seems that in the chicken only BMPR1b (but not BMPR1a) enhances its own expression in response to ligand binding. TGF-beta beads also induce the expression of noggin in the interdigital regions but only with a time delay: noggin expression is detected after 30 hours, while bmpr1b expression becomes detectable in the interdigital region already after 10 hours. TGF-beta thus most likely induces noggin expression indirectly by inducing bmpr1b expression \cite{Merino:1998ha}. 

There are important differences between mouse and chicken limb development as there are differences between forelimb and hindlimb development. We will therefore mainly focus on mouse forelimb bud development and comment on other limb buds as appropriate. In the mouse both Bmpr1a and BmpR1b expression is elevated in the precartilaginous condensations and Bmpr1a and Bmpr1b can compensate for each other \cite{Yoon:2005ka, Montero:2008bp}. BMPR1b null mice show normal digit patterning and no differences in Sox9 expression are observed in wild-type and BmprIB-/- limbs up to and including E12.5, indicating that the prechondrogenic limb mesenchyme is specified and is able to form condensations up to this time \cite{Ovchinnikov:2006p49227,Yi:2000vt,Yoon:2005ka}. BmprIA-/- embryos are not viable and the severity of the BmprIA-/- conditional knock-out depends on the stage at which BmprIA-/- is removed. Cre-mediated excision in Col2-Cre mice seems to occur only in forming cartilaginous condensations, while in Prx1-Cre mice, it acts in limb mesenchyme at much earlier stages (95\% active by E.10.5) \cite{Ovchinnikov:2006p49227}. If BMP receptor type IA is removed with Col2-Cre digit condensations can still be observed while in Prx1-Cre mice no digits form in the forelimb \cite{Ovchinnikov:2006p49227}. The mutants have shortened limbs and show almost complete agenesis of the autopod because of reduced cell proliferation \cite{Ovchinnikov:2006p49227}. Moreover, the expression of BMP target genes Msx1/2 and Grem are severely reduced. BMPR1b appears to rescue Msx1/2 expression. However, the expression is then colocalized with the patchy BMPR1B expression \cite{Ovchinnikov:2006p49227}. In the mouse BMPR1a thus appears to be the major transducer of BMP signals in limb condensations. 

In the limb there are three important BMP ligands: BMP2, BMP4, and BMP7 \cite{Robert:2007fq}. BMP2 and BMP4 appear to be  the most important BMP ligands, but also in the BMP2: BMP4 double conditional knock-out some digits still form \cite{Bandyopadhyay:2006p41317}. At E12.5 BMP2 is expressed predominantly in the interdigital webbing, BMP4 is expressed along the whole AER, and BMP7 expression is ubiquitous in the limb bud \cite{Robert:2007fq}. Bmp2 conditional mutants form five digits, conditional inactivation of a gene for BMP4 results in a polydactylous phenotype \cite{Selever:2004p14564}, but in BMP2:BMP4 conditional mutants two posterior digits are missing even though the limb field is broader \cite{Bandyopadhyay:2006p41317}. Bmp7 null mice show an occasional anterior polydactyly \cite{Robert:2007fq}. Importantly, even though BMP4 expression is reduced in the conditional BMP4 knock-out BMP signaling activity is increased \cite{Benazet:2009p2903}. To avoid the redundant role of multiple BMP ligands and receptors  Zeller and co-workers conditionally removed the downstream signaling protein Smad4 (Co-Smad). The mutants still express Sox9 in the autopod but the pattern does not split into digits  \cite{Benazet:2012wg}. Smad4-dependent signaling is therefore necessary for digit formation. 

BMP signaling is embedded in a larger network that most importantly comprises FGFs and SHH in the limb. BMP signaling interferes with FGF-dependent signaling, at least in part by repressing expression of its receptor FGFR1 \cite{Yoon:2006hj}. FGF-loaded beads in turn repress BMPR1b and Noggin expression in chicken limb buds \cite{Merino:1998ha}. In mouse limb buds FGFs repress the expression of the BMP antagonists Gremlin \cite{Verheyden:2008p13313} and enhance the expression of SHH  in the ZPA, which in turn enhances Gremlin expression. SHH induces the expression of BMP-2 and BMP-7 in chicken limb buds, and with a delay that of BRK2 (BMPR1b) \cite{Kawakami:1996vo}. Without SHH digits cannot form unless expression of Gli3 is removed as well \cite{Litingtung:2002p13571}. In the absence of SHH signaling Gli3 forms the Gli3 repressor which prevents the expression of Gremlin and many other genes. However, in the absence of both SHH and Gli3 more digits appear than in the wildtype. Neither SHH nor Gli3 are therefore necessary for digit formation. Both Fgf4 and SHH expression terminate around E12 \cite{Towers:2009p18309} but Fgf8 continues to be expressed in the AER. 

Given the complex regulatory interactions it is difficult to understand and predict the regulatory outcome by verbal reasoning. We therefore sought to build and analyse a computational model of these interactions. In light of the presence of digits without Shh/Gli3, and the absence of digits in BMPR1A  and SMAD4 conditional mutants we focus on BMP-dependent signaling to explain the observed patterning of Sox9 in the limb bud. We find that the interactions between BMP and BMP receptor give rise to a Schnakenberg-type Turing mechanism. When we solve the model on a realistic domain we obtain patterns that are similar to those observed in experiments with wildtype and mutant mice. We propose that the observed patterning during limb bud development may be the consequence of a Turing type patterning mechanism that arises from the interactions between BMPs and their receptor.

\section*{Results}
\subsection*{Model}
The core signaling pathways that regulate limb patterning have been well characterized and comprise, SHH, GLI, Gremlin, BMP, FGF, and their receptors \cite{Zeller:2010p41336,Benazet:2009p41334}. Digit patterning is still observed in the absence of Gremlin \cite{Zuniga:1999p21732}, Noggin \cite{Brunet:1998p28177}, SHH and GLI3\cite{Litingtung:2002p13571}, but not in the absence of BMPR1A \cite{Ovchinnikov:2006p49227} and Smad4, an essential protein in BMP signaling \cite{Benazet:2012wg}. We therefore neglect all dispensable network elements and focus on BMP signaling to explain digit patterning. We will include FGF as an important modulator and growth factor. Shh-dependent signaling is also an important modulator of BMP signaling in the limb bud and is clearly important to establish the anterior-posterior polarity. Turing patterns have been shown to persist in the presence of such a modulating gradient \cite{Glimm:2012bb}. Accordingly, we focus on the mechanism that enables the emergence of digits in the autopod and we will neglect the detailed effects of Shh and how these enable the establishment of anterior-posterior polarity that results in the formation of a thumb and a pinky. 

BMP (which we denote by $B$) and FGF (denoted by $F$) are secreted proteins and diffuse much faster than the BMP receptor (denoted by $R$) which resides in the plasma membrane. Experiments on the BMP \textit{Drosophila} homologue Dpp suggest that the majority of ligand-bound receptors, $C$, are internalised rapidly and reside mainly inside the cell \cite{Kicheva:2007p3904}. We will therefore ignore diffusion of the receptor-ligand complex. We write $\overline{D}_\mathrm{j}$ ($j = B, R, F$) for the diffusion coefficients with $\overline{D}_\mathrm{R} \ll \overline{D}_\mathrm{B}, \overline{D}_\mathrm{F}$\cite{Kumar2010, Hebert2005}. We write $\overline{D}_{\cdot} \overline{\Delta}{[\cdot]}$ for the diffusion fluxes where $\overline{\Delta} $ denotes the Laplacian operator in Cartesian coordinates, and $[\cdot]$ concentration. The characteristic length of gradients depends both on the speed of diffusion and their removal. In the absence of contrary experimental evidence we will assume the simplest relation, linear decay, at rates $d_{\mathrm{j}} [j]$ for all components (i.e. $j = B, R, C, F$).

BMPs are dimers and one BMP molecule can therefore bind two receptors \cite{Scheufler:1999fv}. The rate of BMP-receptor binding is therefore proportional to $R^2B$, and we use k$_{on}$ and k$_{off}$ as the binding and dissociation rate constants. BMP2 signaling has been shown to reduce BMP2 expression in the limb bud \cite{Bastida:2009p24269} and we therefore make the rate of BMP production negatively dependent on the BMP-receptor complex, $C$, i.e. we write for the BMP production rate $P_B \frac{K_B}{K_B + [\mathrm{C}]}$ and thus obtain for the BMP and BMP-receptor dynamics
\begin{eqnarray} \label{eqB}
\dot{\![\mathrm{B}]} &=& \underbrace{\overline{D}_\mathrm{B} \overline{\Delta} [\mathrm{B}]}_{\text{\ diffusion}} + \underbrace{P_B \frac{K_B}{K_B + [\mathrm{C}]}}_{\text{\ production}} \underbrace{-d_{\mathrm{B}}[\mathrm{B}]}_{\text{\ degradation}} \underbrace{-\, \mathrm{k}_\mathrm{on}[\mathrm{R}]^2[\mathrm{B}] + k_\mathrm{off}[\mathrm{C}]}_{\text{\ complex formation}}  
\end{eqnarray}
\begin{eqnarray} \label{eqC}
\dot{\![\mathrm{C}]} &=& \underbrace{\, \mathrm{k}_\mathrm{on}[\mathrm{R}]^2[\mathrm{B}] - k_\mathrm{off}[\mathrm{C}]}_{\text{\ complex formation}}  \underbrace{-d_{\mathrm{C}}[\mathrm{C}]}_{\text{\ degradation}}. 
\end{eqnarray}
 Signaling of BMP-bound receptors positively regulates receptor production \cite{Merino:1998ha}, and we therefore must make receptor production dependent on the concentration of $C$. In the absence of contrary data we will use the simplest possible relation for the receptor production rate and write $P_R([C]) = p_R + p_C [C]$ where $p_R$ and $p_C$ are constants, 
\begin{eqnarray} \label{eqR}
\dot{\![\mathrm{R}]} &=& \underbrace{\overline{D}_\mathrm{R} \overline{\Delta} [\mathrm{R}]}_{\text{\ diffusion}} +\underbrace{ p_R + p_C([\mathrm{C}])}_{\text{\ production}} \underbrace{-d_{\mathrm{R}}[\mathrm{R}]}_{\text{\ degradation}} \underbrace{-2\, \mathrm{k}_\mathrm{on}[\mathrm{R}]^2[\mathrm{B}] + k_\mathrm{off}[\mathrm{C}]}_{\text{\ complex formation}}. 
\end{eqnarray}
We can simplify this set of equations if we assume that the dynamics of the receptor-ligand complex are fast such that the receptor-ligand complex assumes a quasi steady-state. The concentration of bound receptor, $C$, is then proportional to $R^2B$, i.e.
\begin{eqnarray} \label{eqC_qstst}
[C] \sim \frac{k_{on}}{k_{off} + d_C} [R]^2 [B] = K_C [R]^2 [B]; \qquad K_C =\frac{k_{on}}{k_{off} + d_C}  .
\end{eqnarray}
Equations (\ref{eqB},\ref{eqR}) are sufficient for pattern formation and would reduce to the classical Turing model of Schnakenberg-type if we were to set $p_C = 2d_C$, and $d_B = 0$.

BMP expression and activity is modulated by FGF and Gremlin \cite{Benazet:2009p2903}. BMP induces Gremlin expression and Gremlin then binds and sequesters BMP in an inactive complex \cite{Nissim:2006p27803, Michos:2004p14509}. Gremlin can in principle be secreted, but experimental evidence suggests that BMP4 activation and secretion are also negatively regulated by an intracellular Gremlin-BMP4 interaction \cite{Sun:2006p22658}. We therefore do not consider Gremlin diffusion. Since experiments suggest that also Gremlin is a dimer \cite{Sudo:2004dq} we assume 1:1 binding between BMP and Gremlin at rate $k_{on}' [B] [G]$ and dissociation at rate $k_{off}' [BG]$ where we write $G$ for Gremlin and $BG$ for the BMP-Gremlin complex. If we model BMP-induced expression by a Hill function $ \frac{[B]^n}{ [B]^n + K_{G}^n}$ with Hill constant $K_G$ and Hill coefficient $n$, and assume that dissociation of the complex (at rate $k_{off}'$) is much faster than its degradation then the previous system of equations would need to be expanded to include
\begin{eqnarray} \label{eqGremlin}
\dot{\![\mathrm{B}]} &=& \underbrace{\overline{D}_\mathrm{B} \overline{\Delta} [\mathrm{B}]}_{\text{\ diffusion}} + \underbrace{P_B \frac{K_B}{K_B + [\mathrm{C}]}}_{\text{\ production}}  \underbrace{-d_{\mathrm{B}}[\mathrm{B}]}_{\text{\ degradation}} \underbrace{-\, \mathrm{k}_\mathrm{on}'[\mathrm{R}]^2[\mathrm{B}] + k_\mathrm{off}'[\mathrm{C}]}_{\text{\ complex formation}} -(\underbrace{\, \mathrm{k}_\mathrm{on}'[\mathrm{G}][\mathrm{B}] - k_\mathrm{off}'[\mathrm{BG}]}_{\text{\ complex formation}} ) \nonumber \\
\dot{\![\mathrm{G}]} &=&  \underbrace{p_g  \frac{[B]^n}{ [B]^n + K_{G}^n} }_{\text{\ production}} \underbrace{-d_{\mathrm{G}}[\mathrm{G}]}_{\text{\ degradation}} \underbrace{-\, \mathrm{k}_\mathrm{on}'[\mathrm{G}][\mathrm{B}] + k_\mathrm{off}'[\mathrm{BG}]}_{\text{\ complex formation}}  \nonumber \\
\dot{\![\mathrm{BG}]} &=& \underbrace{\, \mathrm{k}_\mathrm{on}'[\mathrm{G}][\mathrm{B}] - k_\mathrm{off}'[\mathrm{BG}]}_{\text{\ complex formation}}.
\end{eqnarray}
Experiments show that Gremlin is induced by BMP already early during limb development \cite{Benazet:2009p2903}. If we assume rapid formation of the BMP-Gremlin complex in comparison with other reactions we can use a quasi-steady state approximation for complex concentration:
\begin{eqnarray} \label{eqBG}
\dot{\![\mathrm{BG}]} &=& \underbrace{\, \mathrm{k}_\mathrm{on}'[\mathrm{G}][\mathrm{B}] - k_\mathrm{off}'[\mathrm{BG}]}_{\text{\ complex formation}} =0.
\end{eqnarray}
The terms for complex formation thus disappear from the equations for Gremlin and BMP and the equation for Gremlin  uncouples from the other equations. Gremlin seems thus not to affect the patterning module directly. We therefore ignore the Gremlin-BMP interaction in this model, and we will show that the observed oligodactily in the Gremlin knock-outs can be explained with the reduced limb bud size. 

Given the importance of FGFs for limb patterning we extended the model to also include FGF signaling. Much as BMP, FGFs can diffuse rapidly  \cite{Yu:2009p22070}. BMP expression is induced by FGF and the parameter $P_B$ in \eqref{eqB} thus becomes a function of FGF, $F$, i.e.   $P_B(F) = p_{b} + p_B^*  \frac{[F]^n}{ [F]^n + K_{BF}^n}  \frac{K_B}{K_B + [\mathrm{C}]}$, where $K_{BF}$ and $n$ and the Hill constant and Hill coefficient respectively. Signaling of BMP-bound receptors, $C \sim R^2B$ \eqref{eqC_qstst}, has been found to both stimulate and inhibit FGF-dependent processes \cite{PajniUnderwood:2007p13328, Duprez:1996wd}. Such inconsistent observations may be the result of a concentration-dependency of signaling, and it has been suggested that high levels of BMP signaling interfere with FGF-dependent processes while lower levels may positively regulate FGF expression in the AER \cite{Dudley:2000uq}. Hence FGF activity is best described as $P_F([C]) = p_F  \frac{[C]^n}{ [C]^n + K_{F1}^n} \frac{K_{F2}^n}{[C]^n + K_{F2}^n}$, where $K_{F1} \ll K_{F2}$ are the Hill constants for the activating and the inhibitory impacts of BMP signaling and $n$ is the Hill coefficient. Here we note that FGF expression is restricted to the apical ectodermal ridge (AER), and $p_F$ is therefore non-zero only on the distal boundary of the domain (dashed part of Fig. \ref{fig:model}C; simulated expression patterns are shown in Fig. \textbf{S1}, \textbf{S3}G-J). We then write for the FGF dynamics
\begin{eqnarray} \label{eq_F}
\dot{\![F]} &=& \underbrace{\overline{D}_F \Delta  [F]}_{\text{\ diffusion}}  +  \underbrace{p_F  \frac{[C]^n}{[C]^n + K_{F1}^n} \frac{K_{F2}^n}{[C]^n + K_{F2}^n}}_{\text{\ production}}- \underbrace{d_F [F]}_{\text{\ degradation}} .
\end{eqnarray}
FGFs repress the expression of the BMP antagonist Gremlin, and enhance SHH expression in the ZPA, which in turn enhances Gremlin expression. However, as discussed above the Gremlin dynamics uncouple from the other equations, while the expression of SHH terminates before the expression pattern of Sox9 has assumed the characteristic digit pattern \cite{Towers:2009p18309,Kawakami:2005p44294,Stricker:2011cm} and digit patterning can still be observed in limb buds when expression of Shh (and Gli3) is removed in knock-outs \cite{Litingtung:2002p13571}. We thus neither include Gremlin nor SHH in the model. \\

\paragraph{Summary} We propose that the observed patterning in BMP signaling is the result of regulatory interactions between BMP, $B$, its receptor $R$, and FGFs, $F$ as summarized graphically in Fig. \ref{fig:model}A.  The effective interactions in this model are graphically summarized in Fig. \ref{fig:model}B. As final set of equations we have
\begin{eqnarray}\label{eqmodel}
\dot{\![\mathrm{B}]} &=& \overline{D}_\mathrm{B} \overline{\Delta} [\mathrm{B}] + p_{b} + p_B^*  \frac{[F]^n}{ [F]^n + K_{BF}^n}  \frac{K_B}{K_B + K_C [R]^2[B]} - d_C K_C [R]^2[B] -d_{\mathrm{B}}[\mathrm{B}] \nonumber\\
\dot{\![R]} &=&\overline{D}_\mathrm{R} \overline{\Delta} [\mathrm{R}]  + p_R  + (p_C-2 d_C) K_C [R]^2 [B] - d_R R \nonumber \\   
\dot{\![F]} &=& \overline{D}_F\Delta  [F] +  p_F  \frac{(K_C [R]^2[B])^n}{(K_C [R]^2[B])^n + K_{F1}^n} \frac{K_{F2}^n}{(K_C [R]^2[B])^n + K_{F2}^n}- d_F [F].
\end{eqnarray}
We note that FGFs are not part of the core patterning mechanism. For large FGF concentrations ($[F] \gg K_{BF}$) the equation for $F$ uncouples from the other two equations and if the constitutive BMP expression $p_B$, $p_B^*$ was sufficiently large patterning could be obtained also without FGFs. Further regulatory interactions (like those with Gremlin) may tune the parameter values.

\paragraph{Parameter Values}
The measurement of the parameter values (i.e. diffusion coefficients, production and degradation rates)  \emph{in vivo} is complicated and has only been carried out in few model systems, but not in the limb \cite{Kicheva2007, Yu2009, Ries2009}. However, our conclusions do not depend on the exact values of parameters, but mainly depend on their relative values as can be seen by non-dimensionalizing the model. To non-dimensionalize the model we need to choose characteristic length and time scales, as well as characteristic concentrations. As characteristic time scale we use $T = 1 / \delta_F$, i.e. $t = T \tau$. As characteristic length scale we chose the characteristic length of the FGF gradient $\lambda = \sqrt{D_F /d_F}$, i.e. $x = \lambda \zeta$.  We non-dimensionalize the FGF concentration with respect to the Hill constant $K_{BF}$, i.e. $F = [F] / K_{BF}$, and the BMP and receptor concentrations with respect to the characteristic concentration  $c_0 =  \left(K_B /  K_C \right)^{1/3}$, i.e. $B = [B]/ c_0$, $R =  [R]/ c_0$. After re-grouping parameters as $\delta_i =  \frac{d_i}{d_F} $ except $\delta_C =  \frac{d_C}{d_F}K_C c_0^2$, $\rho_i = \frac{p_{i} }{d_F c_0}$ except $\rho_F = \frac{p_{i} }{K_{BF} d_F}$, $\nu = \frac{p_C}{d_F} K_C c_0^2$, $\kappa_i = \frac{K_{Fi}}{ c_0^3 K_C}=\frac{K_{Fi}}{ K_B}$, and $D_B = \frac{\overline{D}_B}{\overline{D}_F}$, $D_R = \frac{\overline{D}_R}{\overline{D}_F}$ we obtain
\begin{eqnarray} \label{eq_model_nondim}
\frac{\partial B}{\partial \tau} &=& D_B\Delta_\zeta B + \rho_B +\rho_B^* \frac{F^n}{ 1+ F^n} \frac{1} {1+R^2 B}  - \delta_C R^2 B  -\delta_B B \\
\frac{\partial R}{\partial \tau}&=& D_R\Delta_\zeta R +  \rho_R  + (\nu- 2\delta_C) R^2 B   - \delta_R R    \\
\frac{\partial F}{\partial \tau} &=&\Delta_\zeta F+  \rho_F   \frac{(R^2 B)^n}{ (R^2 B)^n + \kappa_1^n} \frac{\kappa_2^n}{ (R^2 B)^n + \kappa_2^n}- F.\label{EqF_Final}
\end{eqnarray}
The dimensionless model contains five parameters less and the patterning mechanism no longer depends on absolute diffusion and decay constants, but only on the relative diffusion coefficients and the relative decay rates. Similarly, the absolute protein concentrations do not matter but only the relative concentrations (as result from the relative expression  and decay rates) relative to the Hill constant and the effective binding constant $K_C$.

\paragraph{Domain and Boundary conditions}
The equations were solved on a domain whose shape was extracted from limb bud images at E12.5 as well as on a growing domain. The idealized domain is shown in Fig. \ref{fig:model}C. We used zero-flux boundary conditions for $B$ and $R$ as the developing limb does not exchange with the surrounding except at the flank which is ignored here. FGF production in the AER was implemented as a flux boundary condition, i.e. 
\begin{eqnarray}
\vec{n} \cdot \nabla F & = &  \rho_F   \frac{(R^2 B)^n}{ (R^2 B)^n + \kappa_1^n} \frac{\kappa_2^n}{ (R^2 B)^n + \kappa_2^n}\label{EqF_production}
\end{eqnarray}
where $\vec{n}$ refers to the unit normal vector. We also considered alternative implementations with production in a thin layer  (Fig. \textbf{S1}B-E) or with Eq. \ref{EqF_Final} defined on the boundary (Fig. \textbf{S1}F). A thin layer with flux boundary conditions (Fig. \textbf{S1}C) and with Eq. \ref{EqF_Final} defined on the boundary  (Fig. \textbf{S1}C) give similar patterns to those observed with flux boundary condition on the outer boundary (Fig. \textbf{S1}A). A thin layer with either constant (Fig. \textbf{S1}E) or BMP-dependent FGF production (Fig. \textbf{S1}D) yielded overall the same pattern (same number of spots) but the shape of the spots was different. 

\paragraph{Modelling Domain Growth}
Proliferation is not uniform within the limb domain but the velocity field has not been published \cite{Boehm:2010p42421}. We therefore measured the proximal-distal axis of limb buds over developmental time and find that the growth rate is approximately linear (Fig. \ref{fig7}A) and we are therefore using a linear growth law with growth rate $v_g$. We can then write for the proximal-distal length of the limb bud $\overline{L}(t)= \overline{L}_0 + \overline{v}_\mathrm{g} t$ where $\overline{L}_0$ denotes the initial proximal-distal length of the limb bud, and $t$ denotes time.

To conserve mass (rather than concentration) the reaction-diffusion equations for the soluble factors $B$ and $F$, equations \ref{eq_model_nondim} and \ref{EqF_Final} must be expanded to include the advection and dilution terms \cite{HydroBook}, i.e.
\begin{equation}
\frac{\partial [\mathrm{X}]}{\partial t} +\nabla (u [\mathrm{X}])=D_\mathrm{x}\overline{\Delta}[\mathrm{X}]+R([\mathrm{X}])
\end{equation}
where $u$ denotes the growth speed. For homogeneous growth at rate $\bar{v}_\mathrm{g}$ in proximal-distal direction $\bar{y}$ we then have $u = \frac{\dot{L}}{L} \bar{y} = \frac{\bar{v}_\mathrm{g}}{\overline{L}_0 + \overline{v}_\mathrm{g} t} \bar{y} $ such that
\begin{equation}
\frac{\partial [\mathrm{X}]}{\partial t} +\frac{\dot{L}}{L} \left( [\mathrm{X}] + \bar{y} \nabla [\mathrm{X}] \right) = \frac{\partial [\mathrm{X}]}{\partial t} + \frac{\overline{v}_\mathrm{g}}{\overline{L}_0+\overline{v}_\mathrm{g} t} \left([\mathrm{X}]+\overline{y} \frac{\partial [\mathrm{X}]}{\partial \overline{y}} \right) = D_\mathrm{x}\overline{\Delta}[\mathrm{X}]+R([\mathrm{X}]). 
\end{equation}
To model local growth at the limb boundary (Fig. \ref{figS4}) the direction of growth was set to be normal to the boundary and proportional to the local FGF concentration, i.e. $\vec{u} =  v_g F^m \vec{n}$ where $v_g$ is the growth speed and $\vec{n}$ is the unit normal vector on the boundary. The exponent $m$ was set to four, even though we note that similar results can be obtained with $m=2$.

\subsection*{Patterning on the limb domain}
\begin{figure}[h]
\begin{center}
\includegraphics[width=0.9\textwidth]{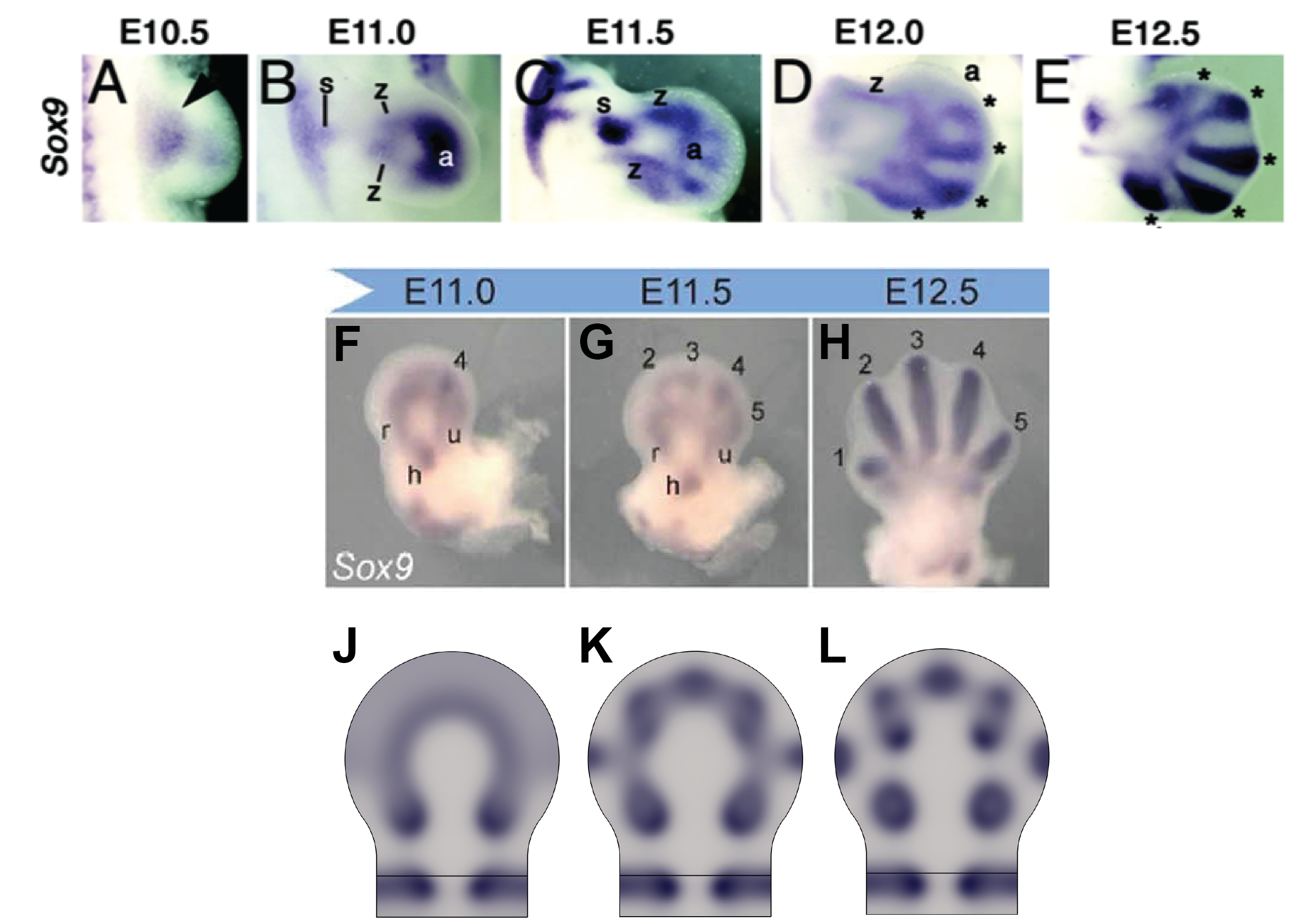}
\caption{\label{fig2} \textbf{The emergence of pattern on the limb bud-shaped domain.} \textbf{(A-H)}   The expression pattern of Sox9 in mouse embryos with developmental time (in embryonic days). Panel (A-E) was reproduced from Fig 1A-E in \cite{Kawakami:2005p44294}; Panel (F-H) was reproduced from Fig 2  in \cite{Stricker:2011cm}. \textbf{(J-L)} The distribution of the $BR^2$ complex at different simulation times \textbf{(J)} $\tau$ = 150 , \textbf{(K)} $\tau$ = 350 (\textbf{(L)} $\tau$ = 750.}
\end{center}
\end{figure}

\begin{figure}[h]
\begin{center}
\includegraphics[width=0.7\textwidth]{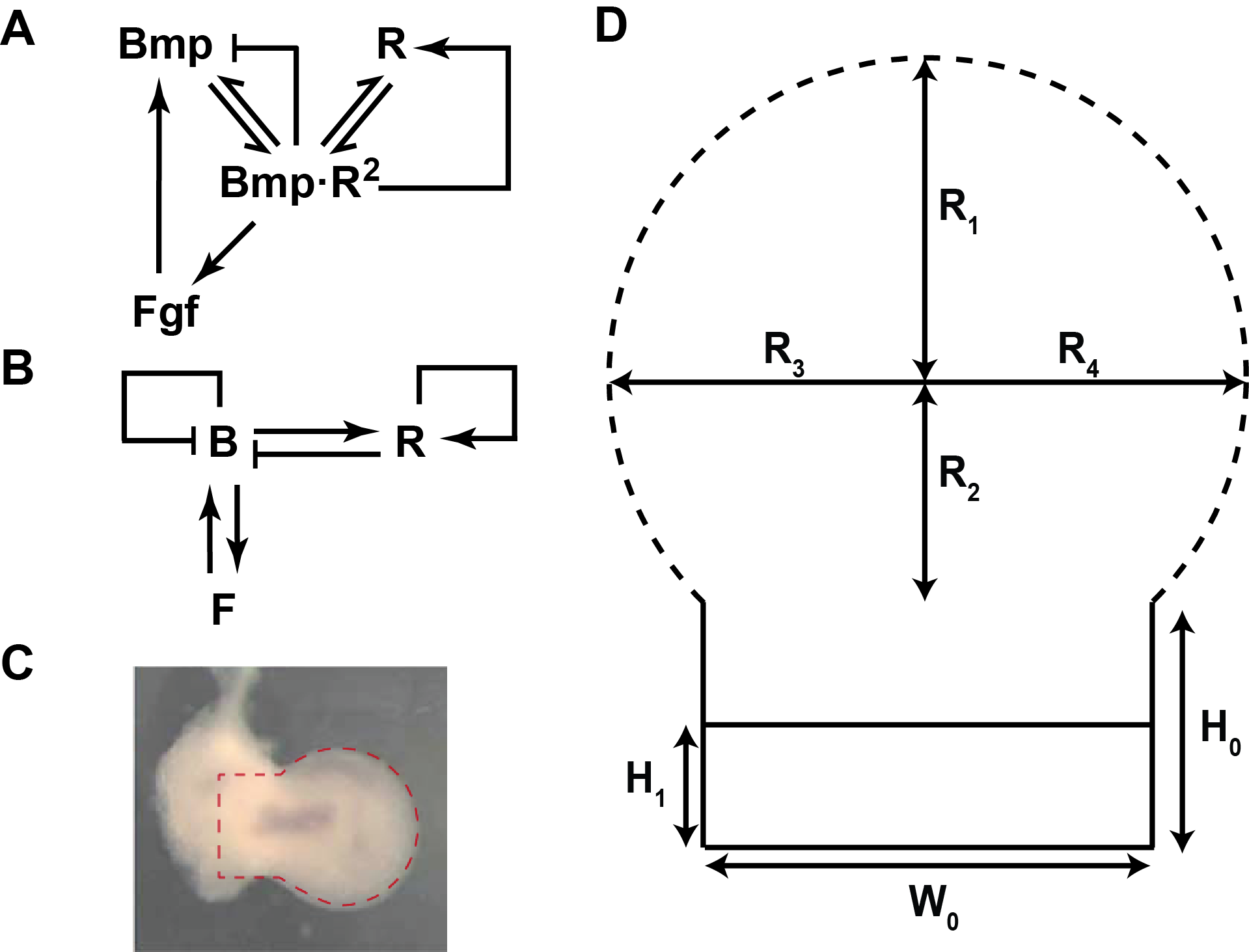}
\caption{\label{fig:model} \textbf{The model.} \textbf{A} The network represents the regulatory interactions that were considered in the model. Thus BMP binds the receptor reversibly to form BMP-receptor complexes. BMP-receptor complexes induce the production of receptors and enhances FGF activity. FGF induces BMP expression.  \textbf{B} The  effective regulatory interactions as captured by Eq \eqref{eq_model_nondim}-\eqref{EqF_Final} . Thus BMP, $B$, has a positive impact on receptors, $R$, and on FGF, $F$, (both via BMP-receptor complexes that are not included here). Receptors are auto-activating (when bound by BMP), while BMP are auto-inhibitory (as they enhance their own decay by receptor binding). FGF and BMP are mutually enhancing each other. For more details see text.  \textbf{(C)} The limb bud shape at E11.5, adopted from  \cite{Stricker:2011cm}.  \textbf{(D)} The computational domain. The values for the different lengths are summarized in Table S1. $R_i$ are the radial axes of the elliptical limb bud. $H_0$ and $W_0$ are the height and width of the stalk. $H_1$ represents the height of the domain in the stalk where the BMP expression is enhanced. The dashed line indicates the AER where AER-FGFs are expressed. For details see text.}
\end{center}
\end{figure}

Since the model is an example of a Schnakenberg Turing-type model \cite{MurrayBook} we expected to find parameter ranges for which we would observe the emergence of patterns. To judge whether such mechanism could yield realistic patterns in a realistic time frame we solved the model on a domain that had been extracted from limb bud images taken at E11.5. Sox9 expression is induced in response to BMP signaling \cite{Pan:2008p44209}, and we were therefore interested in the pattern of the BMP-receptor complex ($C = R^2B$) which will mark the regions of active BMP signaling. This pattern is necessarily distinct from the pattern of BMP expression which remains uniform. In spite of uniform expression of BMP we observe the emergence of pattern in the BMP-receptor signaling domains that resemble the Sox9 pattern observed in experiments (Fig \ref{fig2}J-L). Thus initially (at $\tau=150$) we observe strong BMP signaling (BMP-receptor binding) in the stalk (Fig \ref{fig2}J) (marking the stylopod (s) in the limb bud images in Fig \ref{fig2}B or the humerus (h) in Fig \ref{fig2}F), two distinct spots in the transition zone from stalk to handplate (marking the zeugopod (z) in the limb bud images in Fig \ref{fig2}B or radia (r) and ulna (u) in Fig \ref{fig2}F) and a homogenous expression in the handplate (marking the autopod (a) in the limb bud images in Fig \ref{fig2}B; digit 4 can be discerned already in Fig \ref{fig2}F). Within $\Delta \tau =200$ the pattern in the handplate splits in the simulation (Fig \ref{fig2}K) and distinct condensations can be discerned that mark digits 2-5 in the limb bud images (Fig \ref{fig2}C,G). For the horseshoe pattern to split the BMP expression rate at the proximal end of the domain (rate $\rho_{b2}$) must be 3-fold higher than the BMP expression rate in the "hand plate" (rate $\rho_{b1}$). Without such a local increase in BMP expression the horseshoe pattern would not split.  With time the digit condensations further elongate (Fig \ref{fig2}D,E,H). In the simulations the digit condensations split while elongating (Fig \ref{fig2}L). Elongation of patterns becomes more realistic if we solve the model on a uniformly growing domain (Fig. \ref{fig7}E) or if we allow the domain to expand in direction of the highest FGF concentration (Fig \ref{figS4}). Such deforming growth towards higher FGF concentrations is also observed in cultured limb buds \cite{Niswander:1993p27630,Moon:2000p14615}. To achieve the characteristic digit shape it is important that the FGF concentrates distally from the BMP-receptor spots (Fig \ref{figS4}A-C) as also observed in limb buds (Fig \ref{figS4}D). This is achieved only if BMP-receptor signaling has pre-dominantly a positive impact on FGF expression, i.e. if the BMP concentration is relatively low such that $\kappa_1 \sim R^2 B \ll \kappa_2$. We therefore predict that at the time of digit formation BMP signaling enhances rather than represses FGF8 expression. We should stress that this is necessary only in order to obtain the correct positioning of the FGF8 expression domain. The BMP-receptor pattern described above can be obtained independently of whether BMP regulates FGF signaling positively or negatively. 

In our simulations 5 spots emerge simultaneously in the handplate while in the developing limb digits appear in a sequence (4,2,3/5,1) with digit 1 appearing much later than the other digits. Also the appearance of digit 1 appears to be regulated in a different way from the other digits. Thus digit 1 is still formed in SHH knock-out mutants \cite{Litingtung:2002p13571} while digit 1 is lost in FGF8 knock-out mutants \cite{Moon:2000p14615}. The patterns depend on the specific choice of parameters (Table \ref{tbl:paramValD}), and four instead of five spots could easily be obtained by altering any but the receptor expression rate (Fig. \ref{fig6}A). 
Alternatively the rate of FGF expression $\rho_F$ could be lowered 10-fold on the anterior side initially and then increase over time; such asymmetry in expression pattern is indeed observed in the limb bud. Thus while Fgf8 is expressed uniformly in the AER, expression of Fgf4 appears to be biased to the posterior site of the limb bud. FGF8 is expressed already early during limb bud development and remains expressed throughout the patterning process while Fgf4 expression is first detected around E10.5  and ceases around E12. As a result we expect enhanced FGF production in the posterior part of the limb bud at the onset of Sox9 expression. However, in this paper we will continue to analyse homogenous expression patterns.

\subsection*{Regulatory impact of the parameters}
The patterns depend on the specific choice of parameters (Table \ref{tbl:paramValD}) and since the parameters are difficult to determine accurately in experiments we wondered how sensitive our results would be to variations in parameter values. We find that all model parameters affect the pattern (Fig. \ref{fig6}A). Nonetheless the stable parameter ranges are sufficiently wide that molecular noise should not affect the patterning process. Parameters with particular wide stable ranges include $\delta_B$, the rate of BMP degradation, which can be varied 6-fold without affecting the pattern, as well as the expression rates $\rho_{b2}$, $\rho_R$ and $\rho_F$ for BMP, receptor, and FGF and the threshold for BMP-dependent FGF expression, $\kappa_1$.\\

Since we know the time scale of the process we can relate the non-dimensional kinetic rates to their dimensional counterparts. In our simulations splitting of the pattern in the handplate takes about $\Delta \tau =100$. The processes of one embryonic day (86400 seconds) therefore corresponds to $\Delta \tau \geq 200$ such that $\tau =1$ corresponds to at least $t= T \times \tau =432$ seconds in dimensional time. Accordingly the rate of FGF removal would correspond to $d_F = 1/T \sim  2 \times 10^{-3}$ s$^{-1}$ in dimensional terms which is similar to the experimentally reported value of $10^{-3}$ $s^{-1}$ \cite{Yu2009}. Since in our simulations any change in the rate of FGF removal can be counterbalanced by a change in FGF expression (Fig. \ref{fig6}B) a two-fold lower rate in FGF removal could also be obtained if we reduced the FGF expression rate by about 10-fold. \\

The size of the limb bud changes over time but is about 2 millimeters (mm) at E12.5, the final time of our simulation. Accordingly the length scale is $\lambda = 2$mm$/(H_0+R_1+R_2)$= 2mm/ 19.35 = 0.1 mm and the FGF diffusion coefficient in the model corresponds to 34 $\mu$m$^2$ s$^{-1}$. We further require that BMP diffuses about 2.7-fold faster than FGF which brings the BMP diffusion constant close to the diffusion constant $D \sim 100~\mu m^2 s^{-1}$ that is typically measured for soluble proteins \cite{Yu2009} unless diffusion is impeded by adsorption to the surface \cite{Kicheva2007}. We further require that the BMP receptor diffuses about 100-fold more slowly than the BMP protein. The typical range for diffusional constants of membrane proteins is several orders of magnitude lower than for proteins in solution (i.e. $D=0.1-0.001~\mu m^2 s^{-1}$ \cite{Kumar2010, Hebert2005}). A 100-fold reduced diffusion constant may still overestimate the receptor diffusion constant, in particular because the receptor would then diffuse a distance of 3.5 cell lengths ($l =\sqrt{2D_R t_{1/2}} \simeq 35~ \mu m$) within its receptor half-life time of $t_{1/2}=\ln(2)/d_\mathrm{R}=960~s$ even though receptors are mainly restricted to the surface of single cells. We therefore also tested alternative parameterizations in which the receptor diffusion constant is 400-fold lower than the BMP diffusion constant. Importantly we still obtain qualitative similar, yet smaller and sharper, patterns also with a much lower receptor diffusion constant as long as the FGF and BMP expression are adjusted accordingly. \\

The pattern is also very sensitive to some of the protein expression rates. While there are no reported measurements of protein expression rates and concentrations for the limb bud that these constraints could be related to, there is a large number of mutant phenotypes that can be used for a qualitative comparison. These will be discussed in the following.

\subsubsection*{FGF}
Many different FGFs are expressed during limb development. In the model we consider only those FGFs that are important during digit formation, i.e. FGF4 and FGF8. In mice that are lacking both Fgf4 and Fgf8 function in the forelimb AER limb bud mesenchyme fails to survive and the limb does not develop \cite{Boulet:2004p14537}. FGF4 is expressed only transiently in the limb bud (between E10 and E12 \cite{Lewandoski:2000p14621,Towers:2009p18309}) and removal of only FGF4 does not affect digit patterning \cite{Sun:2000p22667, Moon:2000p14618}. The effect of Fgf8 removal depends on the developmental stage. Inactivating Fgf8 in early limb ectoderm (Msx2-cre;Fgf8 hindlimbs and in RAR-cre;Fgf8 forelimbs) causes a substantial reduction in limb-bud size, a delay in SHH expression, misregulation of Fgf4 expression, and loss of digit 1 (the thumb) \cite{Moon:2000p14615}. In contrast, when functional Fgf8 is transiently expressed, as in Msx2-cre;Fgf8 forelimbs, digit 1 is present and digit 2 or 3 is lost \cite{Lewandoski:2000p14621}. 
If we remove $50\%$ of Fgf expression in our model we observe less separation between the spots, similar to what would be expected in syndactyly (Fig. \ref{fig5}A). To lose a digit we need to reduce FGF expression to less than $10\%$ of its normal levels. In this case we lose the middle digit as is the case in Msx2-cre;Fgf8 forelimbs without causing syndactyly. One explanation for the quantitative differences could be the dominance of FGF8 expression at later stages of limb bud development such that removal of Fgf8 corresponds to removing $90\%$ of FGF activity.  \\

All of the skeletal defects caused by inactivation of Fgf8 can be rescued when Fgf4 is expressed in place of Fgf8  \cite{Lu:2006p14536}. On the protein level FGF4 can thus functionally replace FGF8 in limb skeletal development. An increase in FGF signaling that occurs when the Fgf4 gain-of-function allele is activated in a wild-type limb bud causes formation of a supernumerary posterior digit (postaxial polydactyly), as well as cutaneous syndactyly between all the digits \cite{Lu:2006p14536}. Increasing Fgf expression by 50$\%$ indeed results in supernumerary digits (Fig. \ref{fig5}B). If the additional expression is restricted to the posterior side (where Fgf4 is predominantly  expressed) then the $50\%$ increase results in the posterior placement of one supernumerary digit as well as a merging of spots on the anterior side which would correspond to syndactyly (Fig. \ref{fig5}C).  

The model further predicts that the expression and distribution of FGFs remains homogenous initially, but subsequently forms distinct patches as indeed observed in experiments (Fig. \textbf{S3}G-J). 

\subsubsection*{BMP}
There are several BMPs expressed in the developing limb, including most importantly BMP2, BMP4, and BMP7 \cite{Robert:2007fq}. Much as observed for BMP2 in experiments \cite{Bastida:2009p24269} we predict BMP to be first expressed rather homogenously at the boundary of the domain and to be subsequently expressed most strongly outside the digit domains. In our simulation a  lower BMP production rate leads to digit loss (Fig. \ref{fig5}E) as indeed observed in the BMP2:BMP4 conditional knock-out \cite{Bandyopadhyay:2006p41317}. A $30\%$ increased production rate results in polydactyly in our simulation (Fig. \ref{fig5}F) and ectopically expressed BMP2 indeed induces duplication of digit 2 and bifurcation of digit 3 \cite{Duprez:1996wd}. Moreover, conditional inactivation of the gene for BMP4 results in polydactyly \cite{Selever:2004p14564} and other experiments reveal that even though conditional inactivation of a gene for BMP4 downregulates BMP4 expression, the overall amount of BMP signaling  increases \cite{Benazet:2009p2903}. As the BMP expression rate is further increased in our simulations ($\geq 50\%$ increase) the patterns merge to give rise to polysyndactily (Fig. \ref{fig5}G). Such strong over-expression has not yet been achieved in mutants so that we cannot check this prediction.  

\subsubsection*{BMP receptor} Conditional mutants of BMP receptor type IA lack digits. Since the receptor is an integral part of the proposed Turing mechanism we obtain loss of digits also in the model. Hypomorphic mutants of the BMP receptor have not been described; the model would predict loss of digits (Fig. \textbf{S3}). Mild overexpression of the receptor should lead to a gain of digits while stronger overexpression will lead to a loss of patterning (Fig. \textbf{S3}).  \\

We conclude that the expression rates of BMP and BMP receptor need to be tightly controlled to permit patterning. While FGF is not an essential part of the patterning module FGF enhances BMP production and can thus affect the number of digits.

\subsection*{The impact of domain size and shape}
Turing patterns depend on the size of the computational domain and more modes (patterns) emerge as the size of the domain increases \cite{MurrayBook}. Several mouse mutants have been reported with different shapes and sizes of their limb domain, and larger domains typically correlate with increased digit numbers. Thus GLI3 mutants are polydactylous and the limb domain seems to be widened to 130$\%$  while Gremlin mutants are oligodactylous and the limb domain is shrink to about $60\%$ of its original size  \cite{Zuniga:1999p21732, Litingtung:2002p13571}. Previous theoretical models showed that such a correlation can in principle be explained with Turing models \cite{Miura:2006bd}, but the domain shapes were not based on real limb buds and no quantitative comparison of domain size and digit number was previously carried out. When we shrink the computational domain uniformly to $60\%$ of its original size we observe a decrease to two BMP-receptor ($BR^2$) patches (Fig. \ref{fig3}A,B), much as is observed in experiments \cite{Zuniga:1999p21732}. When we enlarge the computational domain uniformly to $130\%$ of its original size the digit number more than doubles  (Fig. \ref{fig3}A). However, in the GLI3$^{-/-}$ mutant the expansion is observed only along the anterior-posterior axis \cite{Litingtung:2002p13571}. If we increase only the anterior-posterior axis of the limb bud to $130\%$ we observe four additional digits which emerge mainly in the middle of the domain (Fig. \ref{fig3}C,D). A closer analysis of GLI3$^{-/-}$ mutants reveals that limb buds are expanded mainly to the anterior side and that extra digits are concentrated in the anterior domain, while the posterior side is about normal. When we analysed a domain that was expanded by $160\%$ to the anterior and constant on the posterior side we observed three additional spots which now emerge in the anterior side (Fig. \ref{fig3}E,F). This corresponds well to the 8 digits observed in the GLI3-/- null mouse \cite{Litingtung:2002p13571}.

\subsection*{The impact of growth}
\begin{figure}[h]
\begin{center}
\includegraphics[width=0.71\textwidth]{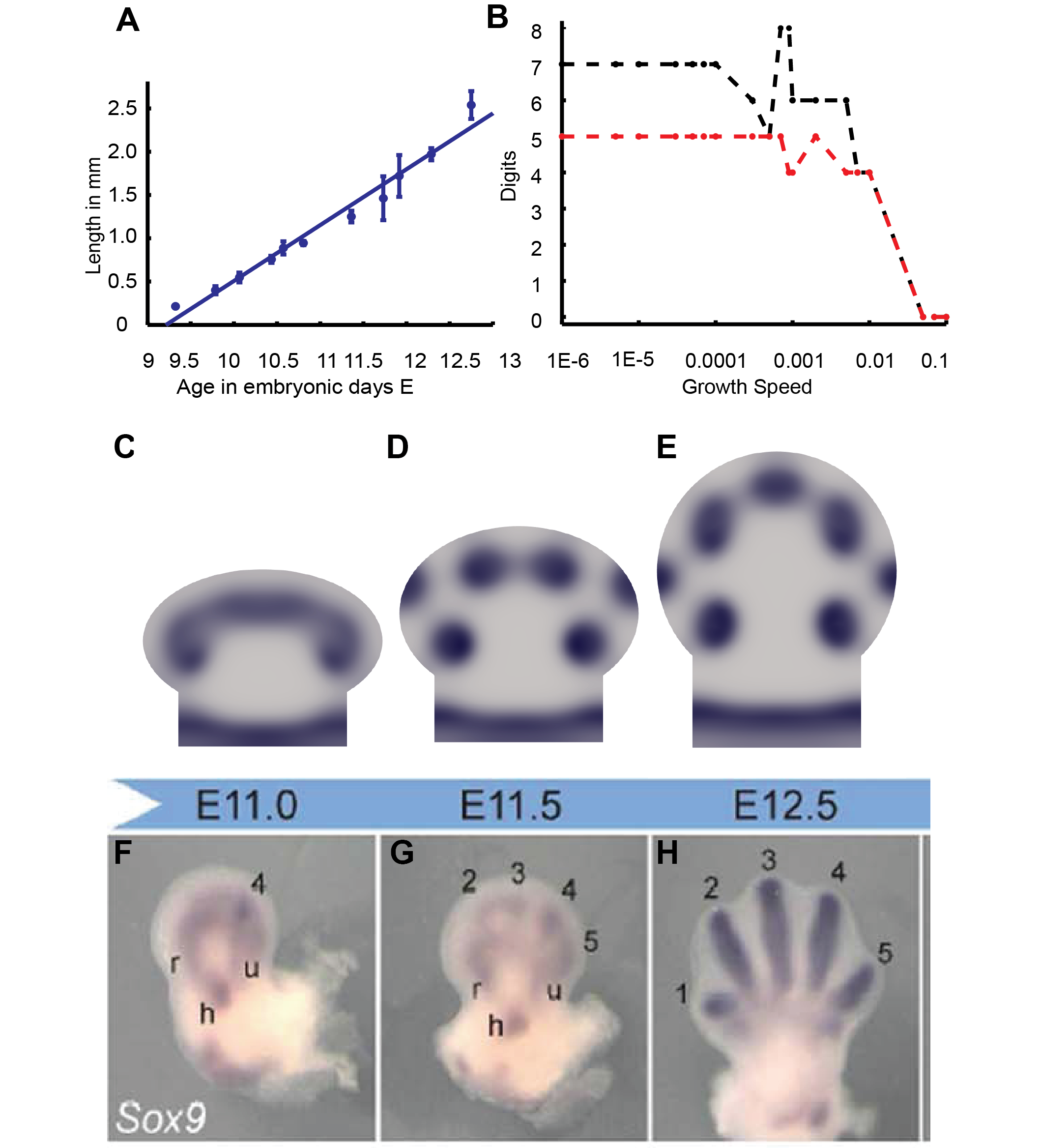}
\caption{\label{fig7} \textbf{Impact of growth on patterning.} \textbf{(A)} The increase in the length of the proximal-distal  mouse limb bud axis with developmental time (in embryonic days) measured from forty limb buds images. Linear growth can be observed at rate of $0.6 \pm 0.1$ mm per day  \textbf{(B)} Impact of the number of spots (digits) on the growth rate. 5 spots (digits) can be obtained over a wider range of growth rates only if we reduce the FGF production rate by 3-fold (${\rho_F}_{growth} \approx {\rho_F}_{static}/3$, red line); ${\rho_F}_{growth} = {\rho_F}_{static}$, black line). \textbf{(C-E)} Snapshots of simulated limb buds at different points. The domain was grown in the proximal-distal axes linearly from 20$\%$ of final size in 8000 timesteps with ${\rho_F}_{growth} \approx {\rho_F}_{static}/3$. To prevent the horse shoe patterning from splitting up into digits early, $\rho_B^*$ had to be decreased from 1.3 to 1 by 0.1 every 2000 time and we set ${\rho}_{B2} = 5 \cdot {\rho}_{B1}$ to ensure the stalk has uniform patterning.  \textbf{(C)} Limb bud at $\tau=4000$ and 60$\%$ of final size.\textbf{(D)}Limb bud at $\tau =5300$ and 73$\%$ of final size.\textbf{(E)}Limb bud at $\tau=8000$ and 100$\%$ of final size.  \textbf{(F-H)}   The expression pattern of Sox9 in mouse embryos with developmental time (in embryonic days) as reproduced from Fig 2  in \cite{Stricker:2011cm}.
}
\end{center}
\end{figure}

\begin{figure}[t!]
\begin{center}
\includegraphics[width=0.75\textwidth]{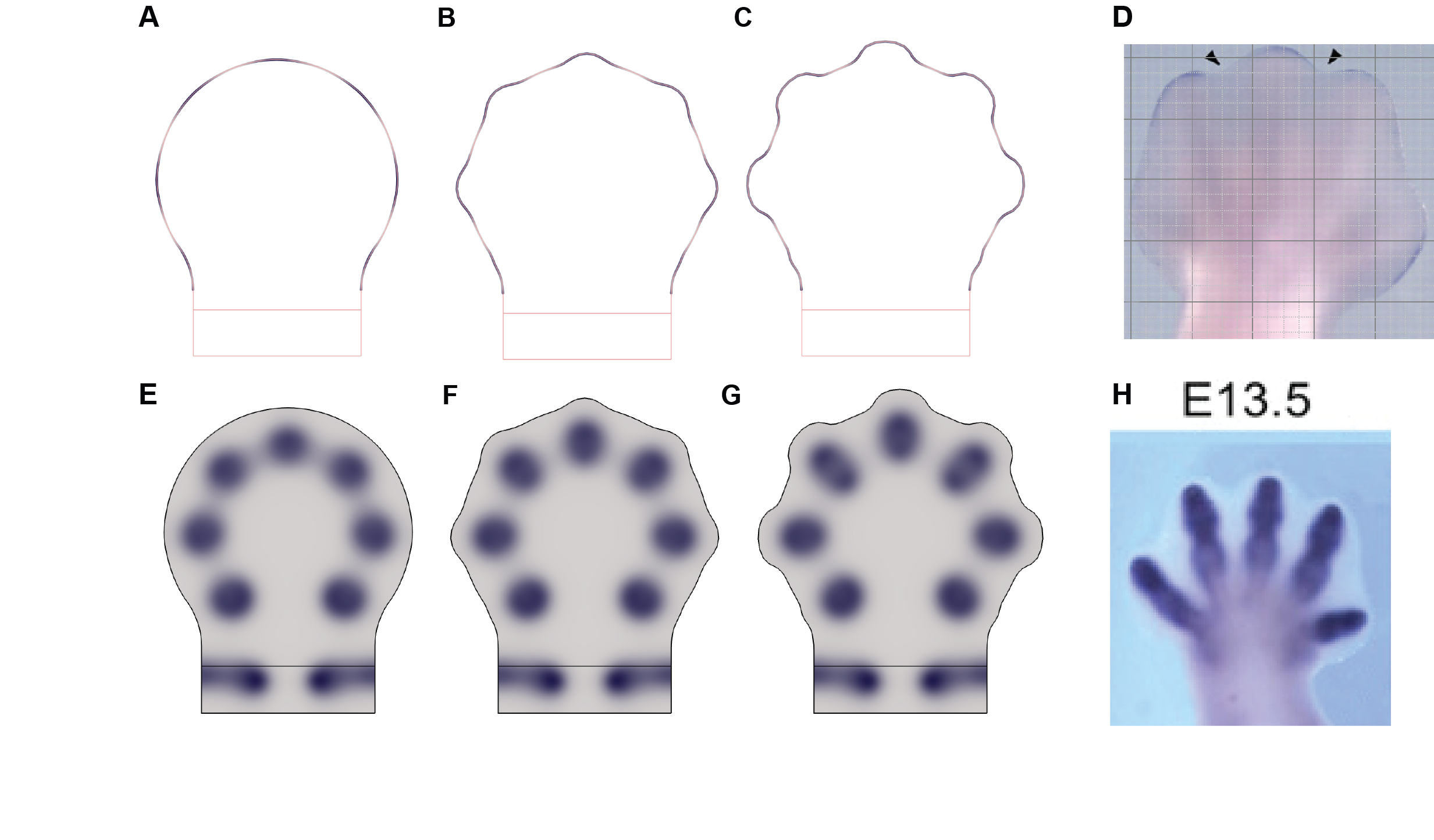}
\caption{\label{figS4} \textbf{Free deformation of the domain according to the local concentration of FGF}.  The FGF patterning  gets split up at the positions of the digits due to the positive feedback of the $BR^2$ complex on FGF. The domain can then be deformed by allowing growth normal to the surface according to the local concentration of FGF. The growth rate is equivalent to $ {[FGF]^4} \times  v_g $. {\textbf(A-C)} Expression of FGF  on the boundary (i.e. $ \rho_B +\rho_B^* \frac{(R^2 B)^n}{ (R^2 B)^n + (K_{B})^n}$) at different simulation times $\tau$ =  500, 1500 and 2500. {\textbf(D)} FGF8 expression pattern in mouse limb at E13.5; adapted from Fig 2C in \cite{Lu:2006p14536}; {\textbf(E-G)} The spatial distribution of the $BR^2$ complex at $\tau$ =  500, 1500 and 2500. {\textbf(H)} Sox9 expression pattern in mouse limb buds at E13.5; adapted from Fig 8M in \cite{Bangs:2011ig}. Note that the rates of receptor and BMP degradation were increased by 10 percent relative to the standard values used.}
\end{center}
\end{figure}

\begin{figure}[h]
\begin{center}
\includegraphics[width=0.5\textwidth]{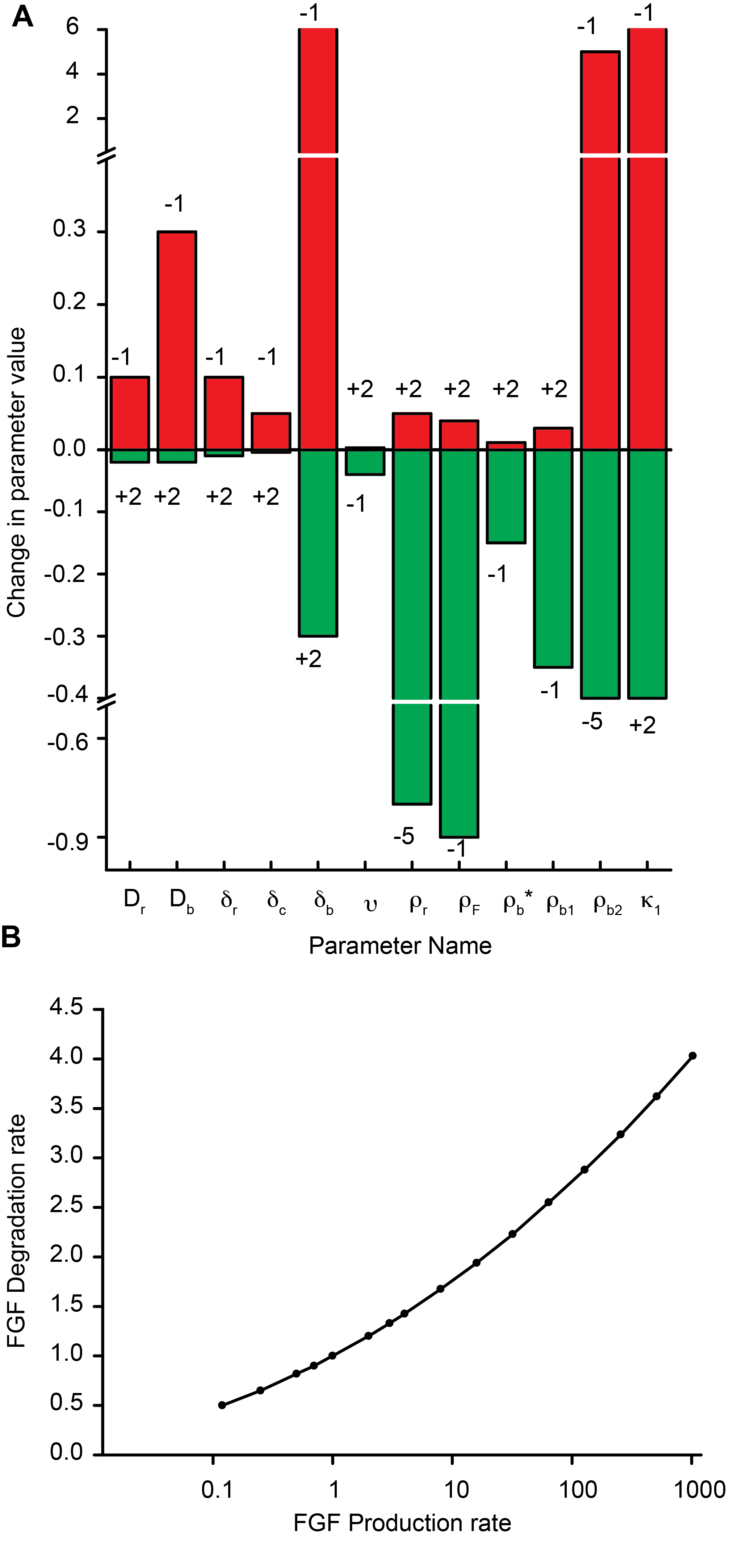}
\caption{\label{fig6}\textbf{Parameter dependency of patterning.}  \textbf{(A)} Local stability analysis. Each parameter as perturbed independently from the reference value given in  Table \ref{tbl:paramValD} as indicated on the vertical axis and the range was recorded for which pattern were qualitatively preserved on a static domain. The numbers above and below the bar indicate how many digits were either gained ($+$) or lost ($-$) as the range was exceeded.  \textbf{(B)} The parameter space for which the pattern is preserved is larger (black line) if parameters are changed together as illustrated for the FGF production and degradation rates. The simulations were analyzed at  $\tau =750$. Please note that the FGF degradation rate determines the time scale of the simulations ($T = 1/d_F$). Compensation with the FGF production rate thus allow us to simulate the processes on different time scales. }
\end{center}
\end{figure}

\begin{figure}[h]
\begin{center}
\includegraphics[width=0.83\textwidth]{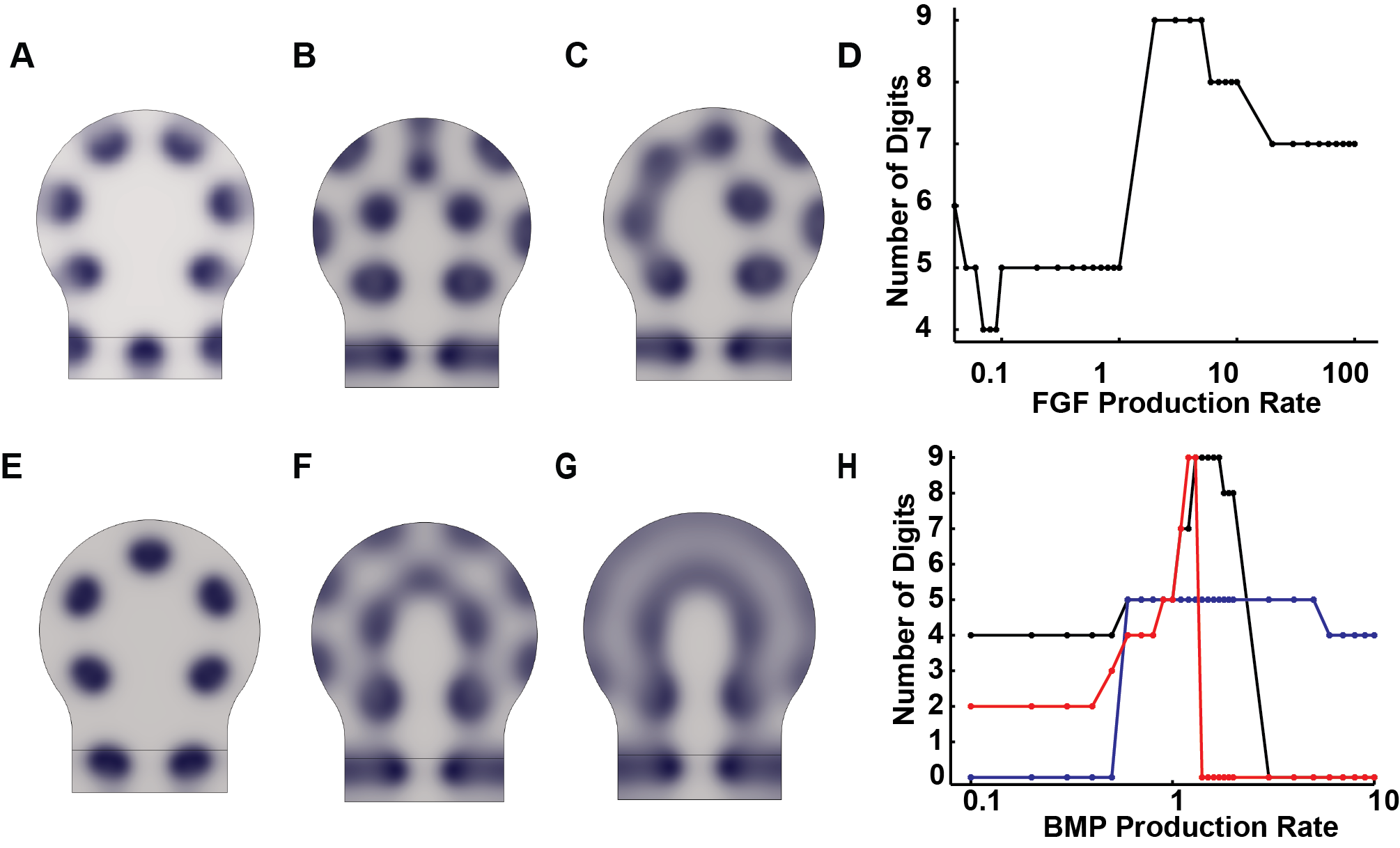}
\caption{\label{fig5} \textbf{The impact of protein production  rates on digit numbers.} \textbf{(A-D)} The impact of changes in the FGF production rate on patterning. \textbf{(A)} A 91$\%$ reduction in the FGF expression rate results in the loss of one pattern (digit). \textbf{(B)} Additional spots emerge as the FGF production rate is increased uniformly by $50\%$. \textbf{(C)} One additional spot emerges on the posterior site (RHS) if FGF expression is enhanced by $50\%$ only on the posterior site (RHS). Spots on the anterior side (LHS) merge as characteristic for syndactyly. \textbf{(D)} The number of digits at different relative FGF production rates (1 corresponds to the reference rate in Table \ref{tbl:paramValD}) for $\tau =750$. Loss of digits at higher FGF production rates is due to presence of stripes; spots can be recovered if simulations are run longer.  \textbf{(E-H)} The impact of changes in the BMP production rate. \textbf{(E)} Digits are lost when the BMP production rate is reduced to 50$\%$. \textbf{(F)} Additional spots emerge as the BMP production rate is increased by $30\%$. \textbf{(G)} The pattern merge (polysyndactily) as the BMP production rate is increased by $50\%$.  \textbf{(H)} The number of digits at different relative BMP production rates (1 corresponds to the reference rate in Table \ref{tbl:paramValD})  for $\tau = 750$. Note that at higher production rates pattern merge and digits are no longer observed.The constant but spatially modulated BMP production rates $\rho_{B1}$, $\rho_{B2}$ (black and blue lines) and the FGF-dependent BMP production rate  $\rho_{B^*}$ (red line) have different effects on the patterning. The spatially modulated BMP $\rho_{B2}$  in the stalk has little impact at larger values because most BMP then binds to receptors in the stalk. The spot(s) at the proximal end of the stalk and the two spots appearing at the transition from the stalk to the circular domain were preserved under all conditions and thus not counted. }
\end{center}
\end{figure}

\begin{figure}[h]
\begin{center}
\includegraphics[width=0.5\textwidth]{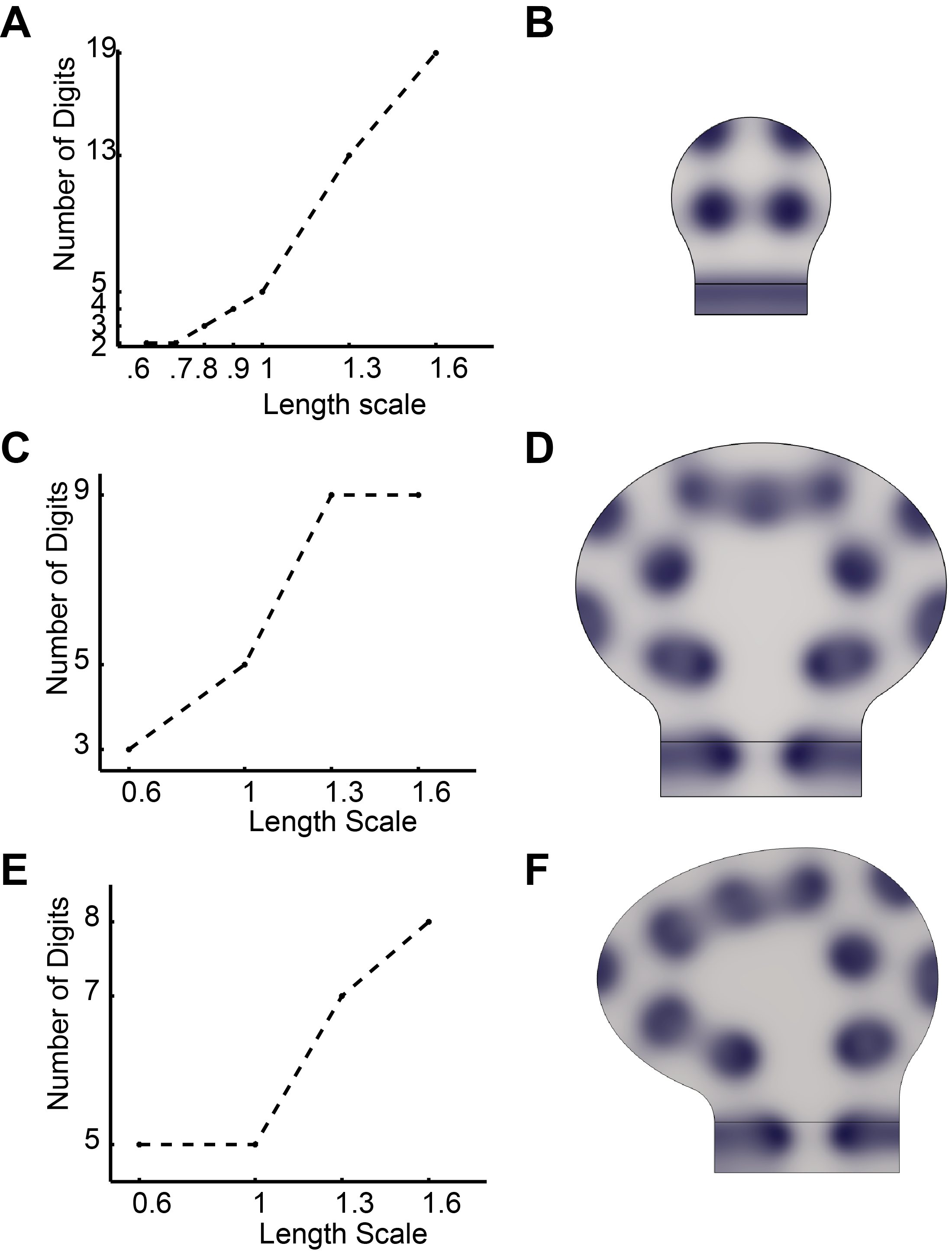}
\caption{\label{fig3} \textbf{Impact of domain size on patterning.}  \textbf{(A)} Number of spots (digits) on a domain that is changed uniformly in its size. \textbf{(B)} Patterning on a domain that is uniformly shrunk to $60\%$ of its normal size. \textbf{(C)} Number of spots (digits) on a domain where the length of the posterior and anterior axes (R3 and R4) are changed as indicated. \textbf{(D)} Patterning on a domain where the AP axis (R3 and R4) is expanded to $130\%$ of its normal size. \textbf{(E)} Number of spots (digits) on a domain where the length of the anterior axis (R3) is changed as indicated. \textbf{(F)} Patterning on a domain where the anterior axis (R3) is expanded to $160\%$ of its normal size. Note: The spot(s) at the proximal end of the stalk and the two spots appearing at the transition from the stalk to the circular domain were preserved under all conditions and are thus not counted in panels A,C,E. }
\end{center}
\end{figure}

The limb bud grows while patterns emerge. Given the importance of domain size and shape we wondered how growth would affect the patterning process. We measured images of limb buds between E9 and E13 and noticed an almost linear growth in proximal-distal direction (Fig. \ref{fig7}A). Limb buds increase in length by about 5-fold within the first four days of development. When we solved the model on a linearly, uniformly growing domain we obtain similar patterns as before (Fig. \ref{fig7}C). However, the number of digits strongly depends on the growth speed (Fig. \ref{fig7}B). For low growth rates we can adjust the number of digits with the help of the other parameters. For increasing growth speeds the speed greatly affects the patterning process and we both lose and gain digits (Fig. \ref{fig7}B). As the growth speed further increases patterns merge and eventually all digits are lost (Fig. \ref{fig7}B). From Figure \ref{fig7}A we can estimate the growth speed as about 7.7 nanometers per second. In dimensionless terms ($T=432$ sec and $\lambda=0.1$ mm) this corresponds to a growth speed of about $v_g  =0.03$ which is too large to permit patterning (Fig. \ref{fig7}B). This may be the reason why there is a the complex pre-patterning process that we do not consider in this model and which starts already around E9.5  \cite{Benazet:2009p41334,Zeller:2009p24283}. The complex regulatory interactions prior to the emergence of digits adjust the expression domains of BMP and FGFs in the limb bud. Such embedding of a Turing mechanism into a sophisticated regulatory network may be important to permit the patterning process to proceed on a rapidly growing domain. Further experiments combined with more detailed modelling also of the other parts of the regulatory network present in the developing limb will, however, be required to either confirm a receptor-ligand based Turing mechanism (embedded in a wider network) or to define an alternative mechanism for digit patterning on the rapidly growing limb bud.

\section*{Discussion}
Limb development has been studied for decades and much is known about the molecular networks that regulate limb development  \cite{Benazet:2009p41334,Zeller:2009p24283}. Based on the available information BMP signaling appears to be the key signaling pathway that regulates the expression of Sox9, the first marker of digit condensations. When modelling BMP signaling we noticed that the regulatory interactions with its receptor constitute a Turing mechanisms. Pattern thus easily emerge. The emerging pattern sensitively depended on the shape and size of the domain, and only when we solved the set of equations on a domain that had been extracted from limb bud images we could reproduce patterns of BMP-receptor activity that resembled the experimentally observed patterns for Sox9. An important test for the suitability of a mathematical model is its consistency of model predictions with a wide range of independent experimental observations. Limb bud development has been studied intensively and a large body of experimental results exists to test the model with. If parameters are chosen appropriately also the phenotype of available mutants can be reproduced. The mechanism is robust to small changes in the parameters as may arise from molecular noise.

Interestingly, while we achieve good correspondence to available experimental data on the static domain, formation of a Turing pattern fails on a domain that grows as fast as measured in the embryo. We propose that the early pre-patterning that involves also SHH/Ptch/Gli signaling as well as the many other regulatory interactions that we ignore in this simple model are in place to support the patterning mechanism on a rapidly growing domain. We are currently developing a simulation for these processes to better understand the impact of these earlier regulatory processes. 

Many models have previously been proposed to explain the emergence of digits, and Turing models in particular have been found previously to be sufficiently flexible to recapitulate the various patterning aspects  of digit emergence \cite{Newman:1979vo, Maini:1991us, Newman:2008p12951}. So far, however, it had been difficult to link the components of a potential Turing mechanism to the molecular constituents. We propose that the Turing pattern in the limb arises from a ligand-receptor interaction. For this to work the ligand needs to be a dimer as is the case for BMPs  \cite{Scheufler:1999fv}. Moreover, the receptor needs to diffuse, yet at a much lower speed than the ligands as is indeed the case. We notice that the diffusion constant that we are using in this simulation is on the high end of what is plausible; lower receptor diffusion constants can be accommodated but the pattern is then sharper and smaller. Again it is possible and plausible that the additional regulatory interactions in the limb bud help to expand the pattern. 

We have previously proposed a mechanism for branch point and branch mode selection in the developing lung \cite{Menshykau:2012kg}. Interestingly, also here receptor-ligand interactions gave rise to a Turing pattern that could recapitulate the different observed branching patterns in the developing lung for physiological parameters and a similar mechanism can be identified also for the developing kidney (Menshykau et al, submitted). While all three systems are based on different signaling proteins (i.e. BMPs in the limb, SHH in the lung, and GDNF in the ureter) they may all exploit the same mechanism to pattern an embryonic field in that these proteins are all multimers that interact with their receptors in a way that induces receptor expression. Interestingly, while BMP can affect its own expression either in a positive or negative feedback via FGFs  \cite{Liu:2005fr, PajniUnderwood:2007p13328}, SHH in the lung clearly engages in a negative feedback via FGF10, while GDNF in the ureter engages in a positive feedback via Wnt signaling \cite{Affolter:2009p25219}. We can obtain the same pattern with a positive and a negative feedback in the limb, lung, and ureter. However, we notice that the possible parameter space is wider in case of a positive feedback. It will be interesting to model further embryonic self-organizing systems to see whether a receptor-ligand interaction may be a general paradigm to enable Turing pattern to emerge in developmental systems. 

Further advancements in our understanding of limb development will require the development of three-dimensional models and the inclusion of more signaling factors. The parameterization and validation of such models will require new experimental data that also reveal the three dimensional dynamics. Such information can now be acquired with the help of optical projection tomographs \cite{Sharpe2002}. Further advances in experimental techniques can thus be expected to provide exciting new insights into the regulatory processes of digit formation during limb development. 


\section*{Methods}
 
\subsection*{Numerical Solution of PDEs}
The PDEs were solved with finite element methods as implemented in COMSOL Multiphysics 4.1 and 4.2. COMSOL Multiphysics is a well-established software package and several studies confirm that COMSOL provides accurate solutions to reaction-diffusion equations both on constant\cite{cutress2010} and growing two-dimensional domains\cite{Carin2006, Thummler2007, Weddemann2008}. Both mesh and the time step were refined until further refinement no longer resulted in noticeable improvements as judged by eye. The simulations were optimised for computational efficiency as described in \cite{Germann2011}.


\begin{thebibliography}{100}
\expandafter\ifx\csname url\endcsname\relax
  \def\url#1{\texttt{#1}}\fi
\expandafter\ifx\csname urlprefix\endcsname\relax\def\urlprefix{URL }\fi
\providecommand{\bibinfo}[2]{#2}
\providecommand{\eprint}[2][]{\url{#2}}

\bibitem{Benazet:2009p41334}
\bibinfo{author}{B{\'e}nazet, J.-D.} \& \bibinfo{author}{Zeller, R.}
\newblock \bibinfo{title}{{Vertebrate limb development: moving from classical
  morphogen gradients to an integrated 4-dimensional patterning system.}}
\newblock \emph{\bibinfo{journal}{Cold Spring Harbor perspectives in biology}}
  \textbf{\bibinfo{volume}{1}}, \bibinfo{pages}{a001339}
  (\bibinfo{year}{2009}).

\bibitem{Zeller:2009p24283}
\bibinfo{author}{Zeller, R.}, \bibinfo{author}{L{\'o}pez-R{\'\i}os, J.} \&
  \bibinfo{author}{Zuniga, A.}
\newblock \bibinfo{title}{{Vertebrate limb bud development: moving towards
  integrative analysis of organogenesis}}.
\newblock \emph{\bibinfo{journal}{Nat Rev Genet}}
  \textbf{\bibinfo{volume}{10}}, \bibinfo{pages}{845--858}
  (\bibinfo{year}{2009}).

\bibitem{Tickle:1981p21603}
\bibinfo{author}{Tickle, C.}
\newblock \bibinfo{title}{{The number of polarizing region cells required to
  specify additional digits in the developing chick wing.}}
\newblock \emph{\bibinfo{journal}{Nature}} \textbf{\bibinfo{volume}{289}},
  \bibinfo{pages}{295--298} (\bibinfo{year}{1981}).

\bibitem{Wolpert:1969p21589}
\bibinfo{author}{Wolpert, L.}
\newblock \bibinfo{title}{{Positional information and the spatial pattern of
  cellular differentiation}}.
\newblock \emph{\bibinfo{journal}{Journal of theoretical biology}}
  \textbf{\bibinfo{volume}{25}}, \bibinfo{pages}{1--47} (\bibinfo{year}{1969}).

\bibitem{Riddle:1993p21612}
\bibinfo{author}{Riddle, R.~D.}, \bibinfo{author}{Johnson, R.~L.},
  \bibinfo{author}{Laufer, E.} \& \bibinfo{author}{Tabin, C.}
\newblock \bibinfo{title}{{Sonic hedgehog mediates the polarizing activity of
  the ZPA}}.
\newblock \emph{\bibinfo{journal}{Cell}} \textbf{\bibinfo{volume}{75}},
  \bibinfo{pages}{1401--1416} (\bibinfo{year}{1993}).

\bibitem{LopezMartinez:1995p21648}
\bibinfo{author}{L{\'o}pez-Mart{\'\i}nez, A.} \emph{et~al.}
\newblock \bibinfo{title}{{Limb-patterning activity and restricted posterior
  localization of the amino-terminal product of Sonic hedgehog cleavage}}.
\newblock \emph{\bibinfo{journal}{Current biology : CB}}
  \textbf{\bibinfo{volume}{5}}, \bibinfo{pages}{791--796}
  (\bibinfo{year}{1995}).

\bibitem{Harfe:2004p18349}
\bibinfo{author}{Harfe, B.~D.} \emph{et~al.}
\newblock \bibinfo{title}{{Evidence for an expansion-based temporal Shh
  gradient in specifying vertebrate digit identities}}.
\newblock \emph{\bibinfo{journal}{Cell}} \textbf{\bibinfo{volume}{118}},
  \bibinfo{pages}{517--528} (\bibinfo{year}{2004}).

\bibitem{Kraus:2001p18579}
\bibinfo{author}{Kraus, P.}, \bibinfo{author}{Fraidenraich, D.} \&
  \bibinfo{author}{Loomis, C.~A.}
\newblock \bibinfo{title}{{Some distal limb structures develop in mice lacking
  Sonic hedgehog signaling}}.
\newblock \emph{\bibinfo{journal}{Mechanisms of development}}
  \textbf{\bibinfo{volume}{100}}, \bibinfo{pages}{45--58}
  (\bibinfo{year}{2001}).

\bibitem{Buscher:1997p13518}
\bibinfo{author}{B{\"u}scher, D.}, \bibinfo{author}{Bosse, B.},
  \bibinfo{author}{Heymer, J.} \& \bibinfo{author}{R{\"u}ther, U.}
\newblock \bibinfo{title}{{Evidence for genetic control of Sonic hedgehog by
  Gli3 in mouse limb development}}.
\newblock \emph{\bibinfo{journal}{Mechanisms of development}}
  \textbf{\bibinfo{volume}{62}}, \bibinfo{pages}{175--182}
  (\bibinfo{year}{1997}).

\bibitem{Litingtung:2002p13571}
\bibinfo{author}{Litingtung, Y.}, \bibinfo{author}{Dahn, R.~D.},
  \bibinfo{author}{Li, Y.}, \bibinfo{author}{Fallon, J.~F.} \&
  \bibinfo{author}{Chiang, C.}
\newblock \bibinfo{title}{{Shh and Gli3 are dispensable for limb skeleton
  formation but regulate digit number and identity.}}
\newblock \emph{\bibinfo{journal}{Nature}} \textbf{\bibinfo{volume}{418}},
  \bibinfo{pages}{979--983} (\bibinfo{year}{2002}).

\bibitem{Zhu:2008p14377}
\bibinfo{author}{Zhu, J.} \emph{et~al.}
\newblock \bibinfo{title}{{Uncoupling Sonic hedgehog control of pattern and
  expansion of the developing limb bud}}.
\newblock \emph{\bibinfo{journal}{Dev Cell}} \textbf{\bibinfo{volume}{14}},
  \bibinfo{pages}{624--632} (\bibinfo{year}{2008}).

\bibitem{Ahn:2004p14441}
\bibinfo{author}{Ahn, S.} \& \bibinfo{author}{Joyner, A.~L.}
\newblock \bibinfo{title}{{Dynamic changes in the response of cells to positive
  hedgehog signaling during mouse limb patterning}}.
\newblock \emph{\bibinfo{journal}{Cell}} \textbf{\bibinfo{volume}{118}},
  \bibinfo{pages}{505--516} (\bibinfo{year}{2004}).

\bibitem{Ede:1969ua}
\bibinfo{author}{Ede, D.~A.} \& \bibinfo{author}{Law, J.~T.}
\newblock \bibinfo{title}{{Computer simulation of vertebrate limb
  morphogenesis.}}
\newblock \emph{\bibinfo{journal}{Nature}} \textbf{\bibinfo{volume}{221}},
  \bibinfo{pages}{244--248} (\bibinfo{year}{1969}).

\bibitem{Boehm:2010p42421}
\bibinfo{author}{Boehm, B.} \emph{et~al.}
\newblock \bibinfo{title}{{The role of spatially controlled cell proliferation
  in limb bud morphogenesis.}}
\newblock \emph{\bibinfo{journal}{PLoS Biol}} \textbf{\bibinfo{volume}{8}},
  \bibinfo{pages}{e1000420} (\bibinfo{year}{2010}).

\bibitem{Dillon:1999p12911}
\bibinfo{author}{Dillon, R.} \& \bibinfo{author}{Othmer, H.~G.}
\newblock \bibinfo{title}{{A mathematical model for outgrowth and spatial
  patterning of the vertebrate limb bud.}}
\newblock \emph{\bibinfo{journal}{Journal of theoretical biology}}
  \textbf{\bibinfo{volume}{197}}, \bibinfo{pages}{295--330}
  (\bibinfo{year}{1999}).

\bibitem{Morishita:2008p28252}
\bibinfo{author}{Morishita, Y.} \& \bibinfo{author}{Iwasa, Y.}
\newblock \bibinfo{title}{{Growth based morphogenesis of vertebrate limb bud.}}
\newblock \emph{\bibinfo{journal}{Bulletin of mathematical biology}}
  \textbf{\bibinfo{volume}{70}}, \bibinfo{pages}{1957--1978}
  (\bibinfo{year}{2008}).

\bibitem{Hirashima:2008p12949}
\bibinfo{author}{Hirashima, T.}, \bibinfo{author}{Iwasa, Y.} \&
  \bibinfo{author}{Morishita, Y.}
\newblock \bibinfo{title}{{Distance between AER and ZPA is defined by
  feed-forward loop and is stabilized by their feedback loop in vertebrate limb
  bud}}.
\newblock \emph{\bibinfo{journal}{Bulletin of mathematical biology}}
  \textbf{\bibinfo{volume}{70}}, \bibinfo{pages}{438--459}
  (\bibinfo{year}{2008}).

\bibitem{Probst:2011jo}
\bibinfo{author}{Probst, S.} \emph{et~al.}
\newblock \bibinfo{title}{{SHH propagates distal limb bud development by
  enhancing CYP26B1-mediated retinoic acid clearance via AER-FGF signalling.}}
\newblock \emph{\bibinfo{journal}{Development (Cambridge, England)}}
  \textbf{\bibinfo{volume}{138}}, \bibinfo{pages}{1913--1923}
  (\bibinfo{year}{2011}).

\bibitem{Dillon:2003p12910}
\bibinfo{author}{Dillon, R.}, \bibinfo{author}{Gadgil, C.} \&
  \bibinfo{author}{Othmer, H.~G.}
\newblock \bibinfo{title}{{Short- and long-range effects of Sonic hedgehog in
  limb development}}.
\newblock \emph{\bibinfo{journal}{Proc. Natl. Acad. Sci USA}}
  \textbf{\bibinfo{volume}{100}}, \bibinfo{pages}{10152--10157}
  (\bibinfo{year}{2003}).

\bibitem{Limbpatternphysic:2009fw}
\bibinfo{author}{Newman, S.~A.}
\newblock \bibinfo{title}{{Limb pattern, physical mechanisms and morphological
  evolution - an interview with Stuart A. Newman. Interviewed by Chuong,
  Cheng-Ming.}} (\bibinfo{year}{2009}).

\bibitem{Alber:2008p28312}
\bibinfo{author}{Alber, M.} \emph{et~al.}
\newblock \bibinfo{title}{{The morphostatic limit for a model of skeletal
  pattern formation in the vertebrate limb.}}
\newblock \emph{\bibinfo{journal}{Bulletin of mathematical biology}}
  \textbf{\bibinfo{volume}{70}}, \bibinfo{pages}{460--483}
  (\bibinfo{year}{2008}).

\bibitem{Miura:2006bd}
\bibinfo{author}{Miura, T.}, \bibinfo{author}{Shiota, K.},
  \bibinfo{author}{Morriss-Kay, G.} \& \bibinfo{author}{Maini, P.~K.}
\newblock \bibinfo{title}{{Mixed-mode pattern in Doublefoot mutant mouse
  limb--Turing reaction-diffusion model on a growing domain during limb
  development.}}
\newblock \emph{\bibinfo{journal}{Journal of theoretical biology}}
  \textbf{\bibinfo{volume}{240}}, \bibinfo{pages}{562--573}
  (\bibinfo{year}{2006}).

\bibitem{Hentschel:2004p39051}
\bibinfo{author}{Hentschel, H. G.~E.}, \bibinfo{author}{Glimm, T.},
  \bibinfo{author}{Glazier, J.~A.} \& \bibinfo{author}{Newman, S.~A.}
\newblock \bibinfo{title}{{Dynamical mechanisms for skeletal pattern formation
  in the vertebrate limb.}}
\newblock \emph{\bibinfo{journal}{Proc Biol Sci}}
  \textbf{\bibinfo{volume}{271}}, \bibinfo{pages}{1713--1722}
  (\bibinfo{year}{2004}).

\bibitem{Miura:2004jz}
\bibinfo{author}{Miura, T.} \& \bibinfo{author}{Maini, P.~K.}
\newblock \bibinfo{title}{{Speed of pattern appearance in reaction-diffusion
  models: implications in the pattern formation of limb bud mesenchyme cells.}}
\newblock \emph{\bibinfo{journal}{Bulletin of mathematical biology}}
  \textbf{\bibinfo{volume}{66}}, \bibinfo{pages}{627--649}
  (\bibinfo{year}{2004}).

\bibitem{Oster:1985p39183}
\bibinfo{author}{Oster, G.~F.}, \bibinfo{author}{Murray, J.~D.} \&
  \bibinfo{author}{Maini, P.~K.}
\newblock \bibinfo{title}{{A model for chondrogenic condensations in the
  developing limb: the role of extracellular matrix and cell tractions.}}
\newblock \emph{\bibinfo{journal}{Journal of embryology and experimental
  morphology}} \textbf{\bibinfo{volume}{89}}, \bibinfo{pages}{93--112}
  (\bibinfo{year}{1985}).

\bibitem{Oster:1983wf}
\bibinfo{author}{Oster, G.~F.}, \bibinfo{author}{Murray, J.~D.} \&
  \bibinfo{author}{Harris, A.~K.}
\newblock \bibinfo{title}{{Mechanical aspects of mesenchymal morphogenesis.}}
\newblock \emph{\bibinfo{journal}{Journal of embryology and experimental
  morphology}} \textbf{\bibinfo{volume}{78}}, \bibinfo{pages}{83--125}
  (\bibinfo{year}{1983}).

\bibitem{Izaguirre:2004hl}
\bibinfo{author}{Izaguirre, J.~A.} \emph{et~al.}
\newblock \bibinfo{title}{{CompuCell, a multi-model framework for simulation of
  morphogenesis.}}
\newblock \emph{\bibinfo{journal}{Bioinformatics (Oxford, England)}}
  \textbf{\bibinfo{volume}{20}}, \bibinfo{pages}{1129--1137}
  (\bibinfo{year}{2004}).

\bibitem{Chaturvedi:2005jj}
\bibinfo{author}{Chaturvedi, R.} \emph{et~al.}
\newblock \bibinfo{title}{{On multiscale approaches to three-dimensional
  modelling of morphogenesis.}}
\newblock \emph{\bibinfo{journal}{J R Soc Interface}}
  \textbf{\bibinfo{volume}{2}}, \bibinfo{pages}{237--253}
  (\bibinfo{year}{2005}).

\bibitem{Christley:2007p12953}
\bibinfo{author}{Christley, S.}, \bibinfo{author}{Alber, M.~S.} \&
  \bibinfo{author}{Newman, S.~A.}
\newblock \bibinfo{title}{{Patterns of mesenchymal condensation in a
  multiscale, discrete stochastic model.}}
\newblock \emph{\bibinfo{journal}{Plos Computational Biology}}
  \textbf{\bibinfo{volume}{3}}, \bibinfo{pages}{e76} (\bibinfo{year}{2007}).

\bibitem{Maini:1991us}
\bibinfo{author}{Maini, P.~K.} \& \bibinfo{author}{Solursh, M.}
\newblock \bibinfo{title}{{Cellular mechanisms of pattern formation in the
  developing limb.}}
\newblock \emph{\bibinfo{journal}{International review of cytology}}
  \textbf{\bibinfo{volume}{129}}, \bibinfo{pages}{91--133}
  (\bibinfo{year}{1991}).

\bibitem{Newman:2008p12951}
\bibinfo{author}{Newman, S.~A.} \emph{et~al.}
\newblock \bibinfo{title}{{Multiscale models for vertebrate limb development.}}
\newblock \emph{\bibinfo{journal}{Current topics in developmental biology}}
  \textbf{\bibinfo{volume}{81}}, \bibinfo{pages}{311--340}
  (\bibinfo{year}{2008}).

\bibitem{Turing:1952p868}
\bibinfo{author}{Turing, A.}
\newblock \bibinfo{title}{{The chemical basis of morphogenesis}}.
\newblock \emph{\bibinfo{journal}{Phil. Trans. Roy. Soc. Lond}}
  \textbf{\bibinfo{volume}{B237}}, \bibinfo{pages}{37--72}
  (\bibinfo{year}{1952}).

\bibitem{Kondo:2010bx}
\bibinfo{author}{Kondo, S.} \& \bibinfo{author}{Miura, T.}
\newblock \bibinfo{title}{{Reaction-diffusion model as a framework for
  understanding biological pattern formation.}}
\newblock \emph{\bibinfo{journal}{Science}} \textbf{\bibinfo{volume}{329}},
  \bibinfo{pages}{1616--1620} (\bibinfo{year}{2010}).

\bibitem{Newman:1979vo}
\bibinfo{author}{Newman, S.~A.} \& \bibinfo{author}{Frisch, H.~L.}
\newblock \bibinfo{title}{{Dynamics of skeletal pattern formation in developing
  chick limb.}}
\newblock \emph{\bibinfo{journal}{Science}} \textbf{\bibinfo{volume}{205}},
  \bibinfo{pages}{662--668} (\bibinfo{year}{1979}).

\bibitem{Zhu:2010km}
\bibinfo{author}{Zhu, J.}, \bibinfo{author}{Zhang, Y.-T.},
  \bibinfo{author}{Alber, M.~S.} \& \bibinfo{author}{Newman, S.~A.}
\newblock \bibinfo{title}{{Bare bones pattern formation: a core regulatory
  network in varying geometries reproduces major features of vertebrate limb
  development and evolution.}}
\newblock \emph{\bibinfo{journal}{PLoS ONE}} \textbf{\bibinfo{volume}{5}},
  \bibinfo{pages}{e10892} (\bibinfo{year}{2010}).

\bibitem{Spagnoli:2007eg}
\bibinfo{author}{Spagnoli, A.} \emph{et~al.}
\newblock \bibinfo{title}{{TGF-beta signaling is essential for joint
  morphogenesis.}}
\newblock \emph{\bibinfo{journal}{J Cell Biol}} \textbf{\bibinfo{volume}{177}},
  \bibinfo{pages}{1105--1117} (\bibinfo{year}{2007}).

\bibitem{Lewandoski:2000p14621}
\bibinfo{author}{Lewandoski, M.}, \bibinfo{author}{Sun, X.} \&
  \bibinfo{author}{Martin, G.~R.}
\newblock \bibinfo{title}{{Fgf8 signalling from the AER is essential for normal
  limb development.}}
\newblock \emph{\bibinfo{journal}{Nat Genet}} \textbf{\bibinfo{volume}{26}},
  \bibinfo{pages}{460--463} (\bibinfo{year}{2000}).

\bibitem{Moon:2000p14615}
\bibinfo{author}{Moon, A.~M.} \& \bibinfo{author}{Capecchi, M.~R.}
\newblock \bibinfo{title}{{Fgf8 is required for outgrowth and patterning of the
  limbs}}.
\newblock \emph{\bibinfo{journal}{Nat Genet}} \textbf{\bibinfo{volume}{26}},
  \bibinfo{pages}{455--459} (\bibinfo{year}{2000}).

\bibitem{Sun:2000p22667}
\bibinfo{author}{Sun, X.} \emph{et~al.}
\newblock \bibinfo{title}{{Conditional inactivation of Fgf4 reveals complexity
  of signalling during limb bud development}}.
\newblock \emph{\bibinfo{journal}{Nat Genet}} \textbf{\bibinfo{volume}{25}},
  \bibinfo{pages}{83--86} (\bibinfo{year}{2000}).

\bibitem{Brunet:1998p28177}
\bibinfo{author}{Brunet, L.~J.}, \bibinfo{author}{McMahon, J.~A.},
  \bibinfo{author}{McMahon, A.~P.} \& \bibinfo{author}{Harland, R.~M.}
\newblock \bibinfo{title}{{Noggin, cartilage morphogenesis, and joint formation
  in the mammalian skeleton.}}
\newblock \emph{\bibinfo{journal}{Science}} \textbf{\bibinfo{volume}{280}},
  \bibinfo{pages}{1455--1457} (\bibinfo{year}{1998}).

\bibitem{Xu:2010kv}
\bibinfo{author}{Xu, J.}, \bibinfo{author}{Krebs, L.~T.} \&
  \bibinfo{author}{Gridley, T.}
\newblock \bibinfo{title}{{Generation of mice with a conditional null allele of
  the Jagged2 gene.}}
\newblock \emph{\bibinfo{journal}{Genesis (New York, NY : 2000)}}
  \textbf{\bibinfo{volume}{48}}, \bibinfo{pages}{390--393}
  (\bibinfo{year}{2010}).

\bibitem{Nakayama:2004kp}
\bibinfo{author}{Nakayama, N.} \emph{et~al.}
\newblock \bibinfo{title}{{A novel chordin-like BMP inhibitor, CHL2, expressed
  preferentially in chondrocytes of developing cartilage and osteoarthritic
  joint cartilage.}}
\newblock \emph{\bibinfo{journal}{Development (Cambridge, England)}}
  \textbf{\bibinfo{volume}{131}}, \bibinfo{pages}{229--240}
  (\bibinfo{year}{2004}).

\bibitem{Towers:2009p18309}
\bibinfo{author}{Towers, M.} \& \bibinfo{author}{Tickle, C.}
\newblock \bibinfo{title}{{Growing models of vertebrate limb development}}.
\newblock \emph{\bibinfo{journal}{Development (Cambridge, England)}}
  \textbf{\bibinfo{volume}{136}}, \bibinfo{pages}{179--190}
  (\bibinfo{year}{2009}).

\bibitem{Kawakami:2005p44294}
\bibinfo{author}{Kawakami, Y.} \emph{et~al.}
\newblock \bibinfo{title}{{Transcriptional coactivator PGC-1alpha regulates
  chondrogenesis via association with Sox9}}.
\newblock \emph{\bibinfo{journal}{Proc. Natl. Acad. Sci USA}}
  \textbf{\bibinfo{volume}{102}}, \bibinfo{pages}{2414--2419}
  (\bibinfo{year}{2005}).

\bibitem{Stricker:2011cm}
\bibinfo{author}{Stricker, S.} \& \bibinfo{author}{Mundlos, S.}
\newblock \bibinfo{title}{{Mechanisms of digit formation: Human malformation
  syndromes tell the story.}}
\newblock \emph{\bibinfo{journal}{Developmental dynamics : an official
  publication of the American Association of Anatomists}}
  \textbf{\bibinfo{volume}{240}}, \bibinfo{pages}{990--1004}
  (\bibinfo{year}{2011}).

\bibitem{Healy:1999p44253}
\bibinfo{author}{Healy, C.}, \bibinfo{author}{Uwanogho, D.} \&
  \bibinfo{author}{Sharpe, P.~T.}
\newblock \bibinfo{title}{{Regulation and role of Sox9 in cartilage
  formation.}}
\newblock \emph{\bibinfo{journal}{Developmental dynamics : an official
  publication of the American Association of Anatomists}}
  \textbf{\bibinfo{volume}{215}}, \bibinfo{pages}{69--78}
  (\bibinfo{year}{1999}).

\bibitem{Zehentner:2002p44310}
\bibinfo{author}{Zehentner, B.~K.}, \bibinfo{author}{Haussmann, A.} \&
  \bibinfo{author}{Burtscher, H.}
\newblock \bibinfo{title}{{The bone morphogenetic protein antagonist Noggin is
  regulated by Sox9 during endochondral differentiation.}}
\newblock \emph{\bibinfo{journal}{Development, growth {\&} differentiation}}
  \textbf{\bibinfo{volume}{44}}, \bibinfo{pages}{1--9} (\bibinfo{year}{2002}).

\bibitem{Zimmerman:1996vv}
\bibinfo{author}{Zimmerman, L.~B.}, \bibinfo{author}{De~Jes{\'u}s-Escobar,
  J.~M.} \& \bibinfo{author}{Harland, R.~M.}
\newblock \bibinfo{title}{{The Spemann organizer signal noggin binds and
  inactivates bone morphogenetic protein 4.}}
\newblock \emph{\bibinfo{journal}{Cell}} \textbf{\bibinfo{volume}{86}},
  \bibinfo{pages}{599--606} (\bibinfo{year}{1996}).

\bibitem{Pereira:2000vn}
\bibinfo{author}{Pereira, R.~C.}, \bibinfo{author}{Economides, A.~N.} \&
  \bibinfo{author}{Canalis, E.}
\newblock \bibinfo{title}{{Bone morphogenetic proteins induce gremlin, a
  protein that limits their activity in osteoblasts.}}
\newblock \emph{\bibinfo{journal}{Endocrinology}}
  \textbf{\bibinfo{volume}{141}}, \bibinfo{pages}{4558--4563}
  (\bibinfo{year}{2000}).

\bibitem{Weatherbee:2006p41231}
\bibinfo{author}{Weatherbee, S.~D.}, \bibinfo{author}{Behringer, R.~R.},
  \bibinfo{author}{Rasweiler, J.~J.} \& \bibinfo{author}{Niswander, L.~A.}
\newblock \bibinfo{title}{{Interdigital webbing retention in bat wings
  illustrates genetic changes underlying amniote limb diversification.}}
\newblock \emph{\bibinfo{journal}{Proceedings of the National Academy of
  Sciences of the United States of America}} \textbf{\bibinfo{volume}{103}},
  \bibinfo{pages}{15103--15107} (\bibinfo{year}{2006}).

\bibitem{Nissim:2006p27803}
\bibinfo{author}{Nissim, S.}, \bibinfo{author}{Hasso, S.~M.},
  \bibinfo{author}{Fallon, J.~F.} \& \bibinfo{author}{Tabin, C.~J.}
\newblock \bibinfo{title}{{Regulation of Gremlin expression in the posterior
  limb bud.}}
\newblock \emph{\bibinfo{journal}{Developmental Biology}}
  \textbf{\bibinfo{volume}{299}}, \bibinfo{pages}{12--21}
  (\bibinfo{year}{2006}).

\bibitem{Liu:2005fr}
\bibinfo{author}{Liu, W.} \emph{et~al.}
\newblock \bibinfo{title}{{Threshold-specific requirements for Bmp4 in
  mandibular development.}}
\newblock \emph{\bibinfo{journal}{Developmental Biology}}
  \textbf{\bibinfo{volume}{283}}, \bibinfo{pages}{282--293}
  (\bibinfo{year}{2005}).

\bibitem{PajniUnderwood:2007p13328}
\bibinfo{author}{Pajni-Underwood, S.}, \bibinfo{author}{Wilson, C.~P.},
  \bibinfo{author}{Elder, C.}, \bibinfo{author}{Mishina, Y.} \&
  \bibinfo{author}{Lewandoski, M.}
\newblock \bibinfo{title}{{BMP signals control limb bud interdigital programmed
  cell death by regulating FGF signaling.}}
\newblock \emph{\bibinfo{journal}{Development (Cambridge, England)}}
  \textbf{\bibinfo{volume}{134}}, \bibinfo{pages}{2359--2368}
  (\bibinfo{year}{2007}).

\bibitem{tenDijke:1994vv}
\bibinfo{author}{ten Dijke, P.} \emph{et~al.}
\newblock \bibinfo{title}{{Identification of type I receptors for osteogenic
  protein-1 and bone morphogenetic protein-4.}}
\newblock \emph{\bibinfo{journal}{The Journal of biological chemistry}}
  \textbf{\bibinfo{volume}{269}}, \bibinfo{pages}{16985--16988}
  (\bibinfo{year}{1994}).

\bibitem{Yamaji:1994bn}
\bibinfo{author}{Yamaji, N.} \emph{et~al.}
\newblock \bibinfo{title}{{A mammalian serine/threonine kinase receptor
  specifically binds BMP-2 and BMP-4.}}
\newblock \emph{\bibinfo{journal}{Biochemical and biophysical research
  communications}} \textbf{\bibinfo{volume}{205}}, \bibinfo{pages}{1944--1951}
  (\bibinfo{year}{1994}).

\bibitem{Verschueren:1995tm}
\bibinfo{author}{Verschueren, K.} \emph{et~al.}
\newblock \bibinfo{title}{{Expression of type I and type IB receptors for
  activin in midgestation mouse embryos suggests distinct functions in
  organogenesis.}}
\newblock \emph{\bibinfo{journal}{Mechanisms of development}}
  \textbf{\bibinfo{volume}{52}}, \bibinfo{pages}{109--123}
  (\bibinfo{year}{1995}).

\bibitem{Greenwald:2003tj}
\bibinfo{author}{Greenwald, J.} \emph{et~al.}
\newblock \bibinfo{title}{{The BMP7/ActRII extracellular domain complex
  provides new insights into the cooperative nature of receptor assembly.}}
\newblock \emph{\bibinfo{journal}{Mol Cell}} \textbf{\bibinfo{volume}{11}},
  \bibinfo{pages}{605--617} (\bibinfo{year}{2003}).

\bibitem{Baur:2000wb}
\bibinfo{author}{Baur, S.~T.}, \bibinfo{author}{Mai, J.~J.} \&
  \bibinfo{author}{Dymecki, S.~M.}
\newblock \bibinfo{title}{{Combinatorial signaling through BMP receptor IB and
  GDF5: shaping of the distal mouse limb and the genetics of distal limb
  diversity.}}
\newblock \emph{\bibinfo{journal}{Development (Cambridge, England)}}
  \textbf{\bibinfo{volume}{127}}, \bibinfo{pages}{605--619}
  (\bibinfo{year}{2000}).

\bibitem{Montero:2008bp}
\bibinfo{author}{Montero, J.~A.}, \bibinfo{author}{Lorda-Diez, C.~I.},
  \bibinfo{author}{Ga{\~n}an, Y.}, \bibinfo{author}{Macias, D.} \&
  \bibinfo{author}{Hurle, J.~M.}
\newblock \bibinfo{title}{{Activin/TGFbeta and BMP crosstalk determines digit
  chondrogenesis.}}
\newblock \emph{\bibinfo{journal}{Developmental Biology}}
  \textbf{\bibinfo{volume}{321}}, \bibinfo{pages}{343--356}
  (\bibinfo{year}{2008}).

\bibitem{Yoon:2005ka}
\bibinfo{author}{Yoon, B.~S.} \emph{et~al.}
\newblock \bibinfo{title}{{Bmpr1a and Bmpr1b have overlapping functions and are
  essential for chondrogenesis in vivo.}}
\newblock \emph{\bibinfo{journal}{Proceedings of the National Academy of
  Sciences of the United States of America}} \textbf{\bibinfo{volume}{102}},
  \bibinfo{pages}{5062--5067} (\bibinfo{year}{2005}).

\bibitem{Kawakami:1996vo}
\bibinfo{author}{Kawakami, Y.} \emph{et~al.}
\newblock \bibinfo{title}{{BMP signaling during bone pattern determination in
  the developing limb.}}
\newblock \emph{\bibinfo{journal}{Development (Cambridge, England)}}
  \textbf{\bibinfo{volume}{122}}, \bibinfo{pages}{3557--3566}
  (\bibinfo{year}{1996}).

\bibitem{Merino:1998ha}
\bibinfo{author}{Merino, R.} \emph{et~al.}
\newblock \bibinfo{title}{{Morphogenesis of digits in the avian limb is
  controlled by FGFs, TGFbetas, and noggin through BMP signaling.}}
\newblock \emph{\bibinfo{journal}{Developmental Biology}}
  \textbf{\bibinfo{volume}{200}}, \bibinfo{pages}{35--45}
  (\bibinfo{year}{1998}).

\bibitem{Ovchinnikov:2006p49227}
\bibinfo{author}{Ovchinnikov, D.~A.} \emph{et~al.}
\newblock \bibinfo{title}{{BMP receptor type IA in limb bud mesenchyme
  regulates distal outgrowth and patterning}}.
\newblock \emph{\bibinfo{journal}{Developmental Biology}}
  \textbf{\bibinfo{volume}{295}}, \bibinfo{pages}{103--115}
  (\bibinfo{year}{2006}).

\bibitem{Yi:2000vt}
\bibinfo{author}{Yi, S.~E.}, \bibinfo{author}{Daluiski, A.},
  \bibinfo{author}{Pederson, R.}, \bibinfo{author}{Rosen, V.} \&
  \bibinfo{author}{Lyons, K.~M.}
\newblock \bibinfo{title}{{The type I BMP receptor BMPRIB is required for
  chondrogenesis in the mouse limb.}}
\newblock \emph{\bibinfo{journal}{Development (Cambridge, England)}}
  \textbf{\bibinfo{volume}{127}}, \bibinfo{pages}{621--630}
  (\bibinfo{year}{2000}).

\bibitem{Robert:2007fq}
\bibinfo{author}{Robert, B.}
\newblock \bibinfo{title}{{Bone morphogenetic protein signaling in limb
  outgrowth and patterning.}}
\newblock \emph{\bibinfo{journal}{Development, growth {\&} differentiation}}
  \textbf{\bibinfo{volume}{49}}, \bibinfo{pages}{455--468}
  (\bibinfo{year}{2007}).

\bibitem{Bandyopadhyay:2006p41317}
\bibinfo{author}{Bandyopadhyay, A.} \emph{et~al.}
\newblock \bibinfo{title}{{Genetic analysis of the roles of BMP2, BMP4, and
  BMP7 in limb patterning and skeletogenesis.}}
\newblock \emph{\bibinfo{journal}{PLoS genetics}} \textbf{\bibinfo{volume}{2}},
  \bibinfo{pages}{e216} (\bibinfo{year}{2006}).

\bibitem{Selever:2004p14564}
\bibinfo{author}{Selever, J.}, \bibinfo{author}{Liu, W.}, \bibinfo{author}{Lu,
  M.-F.}, \bibinfo{author}{Behringer, R.~R.} \& \bibinfo{author}{Martin, J.~F.}
\newblock \bibinfo{title}{{Bmp4 in limb bud mesoderm regulates digit pattern by
  controlling AER development}}.
\newblock \emph{\bibinfo{journal}{Developmental Biology}}
  \textbf{\bibinfo{volume}{276}}, \bibinfo{pages}{268--279}
  (\bibinfo{year}{2004}).

\bibitem{Benazet:2009p2903}
\bibinfo{author}{B{\'e}nazet, J.-D.} \emph{et~al.}
\newblock \bibinfo{title}{{A self-regulatory system of interlinked signaling
  feedback loops controls mouse limb patterning}}.
\newblock \emph{\bibinfo{journal}{Science}} \textbf{\bibinfo{volume}{323}},
  \bibinfo{pages}{1050--1053} (\bibinfo{year}{2009}).

\bibitem{Benazet:2012wg}
\bibinfo{author}{B{\'e}nazet, J.-D.} \emph{et~al.}
\newblock \bibinfo{title}{{Smad4 is required to induce digit ray primordia and
  to initiate the aggregation and differentiation of chondrogenic progenitors
  in mouse limb buds}}.
\newblock \emph{\bibinfo{journal}{Development (Cambridge, England)}}
  \textbf{\bibinfo{volume}{139}}, \bibinfo{pages}{4250--4260}
  (\bibinfo{year}{2012}).

\bibitem{Yoon:2006hj}
\bibinfo{author}{Yoon, B.~S.} \emph{et~al.}
\newblock \bibinfo{title}{{BMPs regulate multiple aspects of growth-plate
  chondrogenesis through opposing actions on FGF pathways.}}
\newblock \emph{\bibinfo{journal}{Development (Cambridge, England)}}
  \textbf{\bibinfo{volume}{133}}, \bibinfo{pages}{4667--4678}
  (\bibinfo{year}{2006}).

\bibitem{Verheyden:2008p13313}
\bibinfo{author}{Verheyden, J.~M.} \& \bibinfo{author}{Sun, X.}
\newblock \bibinfo{title}{{An Fgf/Gremlin inhibitory feedback loop triggers
  termination of limb bud outgrowth}}.
\newblock \emph{\bibinfo{journal}{Nature}} \textbf{\bibinfo{volume}{454}},
  \bibinfo{pages}{638--641} (\bibinfo{year}{2008}).

\bibitem{Zeller:2010p41336}
\bibinfo{author}{Zeller, R.}
\newblock \bibinfo{title}{{The temporal dynamics of vertebrate limb
  development, teratogenesis and evolution.}}
\newblock \emph{\bibinfo{journal}{Curr Opin Genet Dev}}
  \textbf{\bibinfo{volume}{20}}, \bibinfo{pages}{384--390}
  (\bibinfo{year}{2010}).

\bibitem{Zuniga:1999p21732}
\bibinfo{author}{Z{\'u}{\~n}iga, A.} \& \bibinfo{author}{Zeller, R.}
\newblock \bibinfo{title}{{Gli3 (Xt) and formin (ld) participate in the
  positioning of the polarising region and control of posterior limb-bud
  identity}}.
\newblock \emph{\bibinfo{journal}{Development (Cambridge, England)}}
  \textbf{\bibinfo{volume}{126}}, \bibinfo{pages}{13--21}
  (\bibinfo{year}{1999}).

\bibitem{Glimm:2012bb}
\bibinfo{author}{Glimm, T.}, \bibinfo{author}{Zhang, J.},
  \bibinfo{author}{Shen, Y.-Q.} \& \bibinfo{author}{Newman, S.~A.}
\newblock \bibinfo{title}{{Reaction-diffusion systems and external morphogen
  gradients: the two-dimensional case, with an application to skeletal pattern
  formation.}}
\newblock \emph{\bibinfo{journal}{Bulletin of mathematical biology}}
  \textbf{\bibinfo{volume}{74}}, \bibinfo{pages}{666--687}
  (\bibinfo{year}{2012}).

\bibitem{Kicheva:2007p3904}
\bibinfo{author}{Kicheva, A.} \emph{et~al.}
\newblock \bibinfo{title}{{Kinetics of morphogen gradient formation.}}
\newblock \emph{\bibinfo{journal}{Science}} \textbf{\bibinfo{volume}{315}},
  \bibinfo{pages}{521--525} (\bibinfo{year}{2007}).

\bibitem{Kumar2010}
\bibinfo{author}{Kumar, M.}, \bibinfo{author}{Mommer, M.~S.} \&
  \bibinfo{author}{Sourjik, V.}
\newblock \bibinfo{title}{{Mobility of Cytoplasmic, Membrane, and DNA-Binding
  Proteins in Escherichia coli.}}
\newblock \emph{\bibinfo{journal}{Biophysical journal}}
  \textbf{\bibinfo{volume}{98}}, \bibinfo{pages}{552--559}
  (\bibinfo{year}{2010}).

\bibitem{Hebert2005}
\bibinfo{author}{Hebert, B.}, \bibinfo{author}{Costantino, S.} \&
  \bibinfo{author}{Wiseman, P.}
\newblock \bibinfo{title}{{Spatiotemporal image correlation Spectroscopy
  (STICS) theory, verification, and application to protein velocity mapping in
  living CHO cells.}}
\newblock \emph{\bibinfo{journal}{Biophysical journal}}
  \textbf{\bibinfo{volume}{88}}, \bibinfo{pages}{3601--3614}
  (\bibinfo{year}{2005}).

\bibitem{Scheufler:1999fv}
\bibinfo{author}{Scheufler, C.}, \bibinfo{author}{Sebald, W.} \&
  \bibinfo{author}{H{\"u}lsmeyer, M.}
\newblock \bibinfo{title}{{Crystal structure of human bone morphogenetic
  protein-2 at 2.7 A resolution.}}
\newblock \emph{\bibinfo{journal}{Journal of molecular biology}}
  \textbf{\bibinfo{volume}{287}}, \bibinfo{pages}{103--115}
  (\bibinfo{year}{1999}).

\bibitem{Bastida:2009p24269}
\bibinfo{author}{Bastida, M.~F.}, \bibinfo{author}{Sheth, R.} \&
  \bibinfo{author}{Ros, M.~A.}
\newblock \bibinfo{title}{{A BMP-Shh negative-feedback loop restricts Shh
  expression during limb development}}.
\newblock \emph{\bibinfo{journal}{Development (Cambridge, England)}}
  \textbf{\bibinfo{volume}{136}}, \bibinfo{pages}{3779--3789}
  (\bibinfo{year}{2009}).

\bibitem{Michos:2004p14509}
\bibinfo{author}{Michos, O.} \emph{et~al.}
\newblock \bibinfo{title}{{Gremlin-mediated BMP antagonism induces the
  epithelial-mesenchymal feedback signaling controlling metanephric kidney and
  limb organogenesis.}}
\newblock \emph{\bibinfo{journal}{Development (Cambridge, England)}}
  \textbf{\bibinfo{volume}{131}}, \bibinfo{pages}{3401--3410}
  (\bibinfo{year}{2004}).

\bibitem{Sun:2006p22658}
\bibinfo{author}{Sun, J.} \emph{et~al.}
\newblock \bibinfo{title}{{BMP4 activation and secretion are negatively
  regulated by an intracellular gremlin-BMP4 interaction}}.
\newblock \emph{\bibinfo{journal}{The Journal of biological chemistry}}
  \textbf{\bibinfo{volume}{281}}, \bibinfo{pages}{29349--29356}
  (\bibinfo{year}{2006}).

\bibitem{Sudo:2004dq}
\bibinfo{author}{Sudo, S.}, \bibinfo{author}{Avsian-Kretchmer, O.},
  \bibinfo{author}{Wang, L.~S.} \& \bibinfo{author}{Hsueh, A. J.~W.}
\newblock \bibinfo{title}{{Protein related to DAN and cerberus is a bone
  morphogenetic protein antagonist that participates in ovarian paracrine
  regulation.}}
\newblock \emph{\bibinfo{journal}{The Journal of biological chemistry}}
  \textbf{\bibinfo{volume}{279}}, \bibinfo{pages}{23134--23141}
  (\bibinfo{year}{2004}).

\bibitem{Yu:2009p22070}
\bibinfo{author}{Yu, S.~R.} \emph{et~al.}
\newblock \bibinfo{title}{{Fgf8 morphogen gradient forms by a source-sink
  mechanism with freely diffusing molecules}}.
\newblock \emph{\bibinfo{journal}{Nature}} \textbf{\bibinfo{volume}{461}},
  \bibinfo{pages}{533--536} (\bibinfo{year}{2009}).

\bibitem{Duprez:1996wd}
\bibinfo{author}{Duprez, D.~M.}, \bibinfo{author}{Kostakopoulou, K.},
  \bibinfo{author}{Francis-West, P.~H.}, \bibinfo{author}{Tickle, C.} \&
  \bibinfo{author}{Brickell, P.~M.}
\newblock \bibinfo{title}{{Activation of Fgf-4 and HoxD gene expression by
  BMP-2 expressing cells in the developing chick limb.}}
\newblock \emph{\bibinfo{journal}{Development (Cambridge, England)}}
  \textbf{\bibinfo{volume}{122}}, \bibinfo{pages}{1821--1828}
  (\bibinfo{year}{1996}).

\bibitem{Dudley:2000uq}
\bibinfo{author}{Dudley, A.~T.} \& \bibinfo{author}{Tabin, C.~J.}
\newblock \bibinfo{title}{{Constructive antagonism in limb development.}}
\newblock \emph{\bibinfo{journal}{Curr Opin Genet Dev}}
  \textbf{\bibinfo{volume}{10}}, \bibinfo{pages}{387--392}
  (\bibinfo{year}{2000}).

\bibitem{Kicheva2007}
\bibinfo{author}{Kicheva, A.} \emph{et~al.}
\newblock \bibinfo{title}{Kinetics of morphogen gradient formation.}
\newblock \emph{\bibinfo{journal}{Science}} \textbf{\bibinfo{volume}{315}},
  \bibinfo{pages}{521--525} (\bibinfo{year}{2007}).

\bibitem{Yu2009}
\bibinfo{author}{Yu, S.~R.} \emph{et~al.}
\newblock \bibinfo{title}{Fgf8 morphogen gradient forms by a source-sink
  mechanism with freely diffusing molecules.}
\newblock \emph{\bibinfo{journal}{Nature}} \textbf{\bibinfo{volume}{461}},
  \bibinfo{pages}{533--537} (\bibinfo{year}{2009}).

\bibitem{Ries2009}
\bibinfo{author}{Ries, J.}, \bibinfo{author}{Yu, S.~R.},
  \bibinfo{author}{Burkhardt, M.}, \bibinfo{author}{Brand, M.} \&
  \bibinfo{author}{Schwille, P.}
\newblock \bibinfo{title}{Modular scanning fcs quantifies receptor-ligand
  interactions in living multicellular organisms.}
\newblock \emph{\bibinfo{journal}{Nat Methods}} \textbf{\bibinfo{volume}{6}},
  \bibinfo{pages}{643--U31} (\bibinfo{year}{2009}).

\bibitem{HydroBook}
\bibinfo{author}{R, A.}
\newblock \emph{\bibinfo{title}{{Vectors, Tensors and the Basic Equations of
  Fluid Mechanics.}}} (\bibinfo{publisher}{Dover Publications},
  \bibinfo{year}{1989}).

\bibitem{MurrayBook}
\bibinfo{author}{Murray, J.~D.}
\newblock \emph{\bibinfo{title}{{Mathematical Biology. 3rd edition in 2
  volumes: Mathematical Biology: II. Spatial Models and Biomedical
  Applications.}}} (\bibinfo{publisher}{Springer}, \bibinfo{year}{2003}).

\bibitem{Pan:2008p44209}
\bibinfo{author}{Pan, Q.} \emph{et~al.}
\newblock \bibinfo{title}{{Sox9, a key transcription factor of bone
  morphogenetic protein-2-induced chondrogenesis, is activated through BMP
  pathway and a CCAAT box in the proximal promoter}}.
\newblock \emph{\bibinfo{journal}{Journal of cellular physiology}}
  \textbf{\bibinfo{volume}{217}}, \bibinfo{pages}{228--241}
  (\bibinfo{year}{2008}).

\bibitem{Niswander:1993p27630}
\bibinfo{author}{Niswander, L.} \& \bibinfo{author}{Martin, G.~R.}
\newblock \bibinfo{title}{{FGF-4 and BMP-2 have opposite effects on limb
  growth.}}
\newblock \emph{\bibinfo{journal}{Nature}} \textbf{\bibinfo{volume}{361}},
  \bibinfo{pages}{68--71} (\bibinfo{year}{1993}).

\bibitem{Boulet:2004p14537}
\bibinfo{author}{Boulet, A.~M.}, \bibinfo{author}{Moon, A.~M.},
  \bibinfo{author}{Arenkiel, B.~R.} \& \bibinfo{author}{Capecchi, M.~R.}
\newblock \bibinfo{title}{{The roles of Fgf4 and Fgf8 in limb bud initiation
  and outgrowth}}.
\newblock \emph{\bibinfo{journal}{Developmental Biology}}
  \textbf{\bibinfo{volume}{273}}, \bibinfo{pages}{361--372}
  (\bibinfo{year}{2004}).

\bibitem{Moon:2000p14618}
\bibinfo{author}{Moon, A.~M.}, \bibinfo{author}{Boulet, A.~M.} \&
  \bibinfo{author}{Capecchi, M.~R.}
\newblock \bibinfo{title}{{Normal limb development in conditional mutants of
  Fgf4.}}
\newblock \emph{\bibinfo{journal}{Development (Cambridge, England)}}
  \textbf{\bibinfo{volume}{127}}, \bibinfo{pages}{989--996}
  (\bibinfo{year}{2000}).

\bibitem{Lu:2006p14536}
\bibinfo{author}{Lu, P.}, \bibinfo{author}{Minowada, G.} \&
  \bibinfo{author}{Martin, G.~R.}
\newblock \bibinfo{title}{{Increasing Fgf4 expression in the mouse limb bud
  causes polysyndactyly and rescues the skeletal defects that result from loss
  of Fgf8 function.}}
\newblock \emph{\bibinfo{journal}{Development (Cambridge, England)}}
  \textbf{\bibinfo{volume}{133}}, \bibinfo{pages}{33--42}
  (\bibinfo{year}{2006}).

\bibitem{Menshykau:2012kg}
\bibinfo{author}{Menshykau, D.}, \bibinfo{author}{Kraemer, C.} \&
  \bibinfo{author}{Iber, D.}
\newblock \bibinfo{title}{{Branch Mode Selection during Early Lung
  Development.}}
\newblock \emph{\bibinfo{journal}{Plos Computational Biology}}
  \textbf{\bibinfo{volume}{8}}, \bibinfo{pages}{e1002377}
  (\bibinfo{year}{2012}).

\bibitem{Affolter:2009p25219}
\bibinfo{author}{Affolter, M.}, \bibinfo{author}{Zeller, R.} \&
  \bibinfo{author}{Caussinus, E.}
\newblock \bibinfo{title}{{Tissue remodelling through branching
  morphogenesis.}}
\newblock \emph{\bibinfo{journal}{Nat Rev Mol Cell Biol}}
  \textbf{\bibinfo{volume}{10}}, \bibinfo{pages}{831--842}
  (\bibinfo{year}{2009}).

\bibitem{Sharpe2002}
\bibinfo{author}{Sharpe, J.} \emph{et~al.}
\newblock \bibinfo{title}{{Optical Projection Tomography as a Tool for 3D
  Microscopy and Gene Expression Studies.}}
\newblock \emph{\bibinfo{journal}{Science}} \textbf{\bibinfo{volume}{296}},
  \bibinfo{pages}{541--545} (\bibinfo{year}{2002}).

\bibitem{cutress2010}
\bibinfo{author}{Cutress, I.~J.}, \bibinfo{author}{Dickinson, E. J.~F.} \&
  \bibinfo{author}{Compton, R.~G.}
\newblock \bibinfo{title}{{Analysis of commercial general engineering finite
  element software in electrochemical simulations.}}
\newblock \emph{\bibinfo{journal}{J Electroanal Chem}}
  \textbf{\bibinfo{volume}{638}}, \bibinfo{pages}{76--83}
  (\bibinfo{year}{2010}).

\bibitem{Carin2006}
\bibinfo{author}{Carin, M.}
\newblock \bibinfo{title}{{Numerical Simulation of Moving Boundary Problems
  with the ALE Method: Validation in the Case of a Free Surface and a Moving
  Solidification Front.}}
\newblock \emph{\bibinfo{journal}{Excert from the Proceedings of the COMSOL
  Conference}}  (\bibinfo{year}{2006}).

\bibitem{Thummler2007}
\bibinfo{author}{Thummler, V.} \& \bibinfo{author}{Weddemann, A.}
\newblock \bibinfo{title}{{Computation of Space-Time Patterns via ALE
  Methods.}}
\newblock \emph{\bibinfo{journal}{Excert from the Proceedings of the COMSOL
  Conference}}  (\bibinfo{year}{2007}).

\bibitem{Weddemann2008}
\bibinfo{author}{Weddemann, A.} \& \bibinfo{author}{Thummler, V.}
\newblock \bibinfo{title}{{Stability Analysis of ALE-Methods for
  Advection-Diffusion Problems.}}
\newblock \emph{\bibinfo{journal}{Excert from the Proceedings of the COMSOL
  Conference}}  (\bibinfo{year}{2008}).

\bibitem{Germann2011}
\bibinfo{author}{Germann, P.}, \bibinfo{author}{Menshykau, D.},
  \bibinfo{author}{Tanaka, S.} \& \bibinfo{author}{Iber, D.}
\newblock \bibinfo{title}{{Simulating organogensis in Comsol.}}
\newblock \emph{\bibinfo{journal}{Proceedings of COMSOL Conference 2011}}
  (\bibinfo{year}{2011}).

\bibitem{Bangs:2011ig}
\bibinfo{author}{Bangs, F.} \emph{et~al.}
\newblock \bibinfo{title}{{Generation of mice with functional inactivation of
  talpid3, a gene first identified in chicken.}}
\newblock \emph{\bibinfo{journal}{Development (Cambridge, England)}}
  \textbf{\bibinfo{volume}{138}}, \bibinfo{pages}{3261--3272}
  (\bibinfo{year}{2011}).

\end{thebibliography}

\section*{Acknowledgments}
The authors are grateful for permission from Development to reproduce Fig 1A-E from \cite{Kawakami:2005p44294}, Fig 2  from \cite{Stricker:2011cm}, Fig 2C from \cite{Lu:2006p14536} and Fig 8M from \cite{Bangs:2011ig}, to the Zeller group for sharing unpublished data, to Victoria Nemeth and Marcus Groote for discussions of numerical methods, and to the Iber group for the critical reading of the manuscript. 

\section*{Author Contributions}
DI and CK developed the model. AB, CK, PG, and DM carried out the analysis. DI wrote the manuscript. 

\section*{Additional Information} \subsection*{Competing Financial Interests} The authors declare no competing financial interests.

\clearpage

\section*{Figure Legends}

\clearpage
\newpage

\clearpage
\newpage

\clearpage

\clearpage         

\clearpage
\newpage

\section*{Tables}

\begin{table}[h!]
\centering
\caption{Values of the dimensionless parameters. } \label{tbl:paramValD}
\begin{tabular}{l | l | l | l| l}
\hline
Parameter  & Non-dim value & Description \\
\hline
Production Rates \\
\hline
${\rho_F}$ & 243.1 & max production rate of $F$ \\
${\rho}_R$ & 0.0145 & max production rate of $R$  \\
${\nu}$ & $1.361$ & max $C$-dependent rate of $R$ production\\
${\rho_B^*}$ & $12 \cdot \rho_{R}$ & max $F$-dependent rate of $B$ production\\
${\rho}_{B1}$ & $0.063$ & Constant production rate of $B$ in the "handplate" \\
${\rho}_{B2}$ & $3 \cdot \rho_{B1}$ & Constant production rate of $B$ in the lower stalk (H$_1$)   \\
\hline
Diffusion Coefficients \\
\hline
$D_{B}$ & 2.7 & BMP diffusion  constant \\
$D_{R}$ & $0.01\cdot D_B$ & Receptor diffusion constant   \\
\hline
Decay constants \\
\hline
$\delta_R$ & $0.135$& Receptor decay constant \\
$\delta_{B}$ & $0.15 \cdot \delta_{r}$ & BMP decay constant \\
$\delta_C$ & $\nu /3$& Receptor-ligand complex decay constant \\
\hline
Regulation\\
\hline
$n$ & 2 & Hill coefficient \\
$\kappa_{1}$ & 0.06588 & Hill constant for BMP $\rightarrow$ FGF positive feedback \\
$\kappa_{2}$ & very large & Hill constant for BMP $-|$ FGF negative feedback \\
\hline
Domain Size\\
\hline
$R_{Ref}$ & 7.74 & Reference radius\\
\hline
\end{tabular}
\end{table}

\end{document}


\begin{flushleft}
{\Large
\textbf{Digit patterning during limb development as a result of the BMP-receptor interaction}
}
\\[0.25cm]
Amarendra Badugu$^{1}$, 
Conradin Kraemer$^{1}$, 
Philipp Germann$^{2}$, 
Denis Menshykau$^{1}$, 
Dagmar Iber$^{2,\ast}$
\\
\bf{1} Department for Biosystems Science and Engineering (D-BSSE), ETH Zurich, Basel, Switzerland
\\
\bf{2} Department for Biosystems Science and Engineering (D-BSSE) ETH Zurich and Swiss Institute of Bioinformatics, Basel, Switzerland
$\ast$ E-mail: dagmar.iber@bsse.ethz.c
\end{flushleft}

\clearpage
\newpage
\section*{Supplementary Text}

\subsection*{Linear Stability Analysis of a simplified version of the proposed model}
We carry out a linear stability analysis to check whether the proposed model is indeed of Turing type as described in \cite{MurrayBook}. The boundary conditions in the proposed model are rather complicated and the spatial modulation of boundary conditions and Bmp production introduce further complications (Equation 12 in the main manuscript). Here we present the results of a linear stability analysis for a simplified model with constant FGF production on the entire domain and no spatial modulation for BMP. Most of the techniques developed for linear stability analysis on a growing domain assume exponential growth of the domain (e.g. Madzamuze A 2008 Int J Dyn Sys Dif Eq and Madzamudze A et al 2010 Math Biol);  similarly we will  also assume exponential growth for this linear stability analysis.

The Jacobean $J$ at the steady state in the absence of diffusion can be determined as
\begin{eqnarray}
J &=& \left(
\begin{array}{ccc}
 2 B R (\nu - 2\delta_c) - \delta_r & R^2 (\nu-2 \delta_c)   &   0\\
- 2 B R \delta_c - \frac{2 B F^2 R \rho_{b1} }{(1+F^2) (1+BR^2)^2}  & - \delta_b - R^2 \delta_c - \frac{F^2 R^2 \rho_{b1} }{(1+F^2) (1+BR^2)^2}   & - \frac{2 F^3 \rho_{b1} }{(1+F^2)^2 (1+BR^2)}   + \frac{2 F \rho_{b1} }{(1+F^2) (1+BR^2)}   \\
0   & 0   & - \delta_f   
\end{array}
\right) \nonumber \\
&=& 
\left(
\begin{array}{ccc}
39.8 & 172 & 0 \\
-97.7 & -194 & 9.67 \\
0   & 0   & -374
\end{array}
\right)
\end{eqnarray}

We find that for the parameters used in the manuscript all eigenvalues are negative (i.e. $tr(J) < 0$; $\det(J) <0$) in the absence of diffusion, and therefore the steady state is linearly stable. Here we note that the FGF production rate that we had to keep constant was varied in a broad range and for the entire range no positive eigenvalues were observed in the absence of diffusion.

In the presence of diffusion on a constant domain the Jacobean reads as
\begin{eqnarray}
J &=& \left(
\begin{array}{ccc}
 -D_r k^2 + 2 B R (\nu - 2\delta_c) - \delta_r & R^2 (\nu-2 \delta_c)   &   0\\
- 2 B R \delta_c - \frac{2 B F^2 R \rho_{b1} }{(1+F^2) (1+BR^2)^2}  & -D_b k^2 - \delta_b - R^2 \delta_c - \frac{F^2 R^2 \rho_{b1} }{(1+F^2) (1+BR^2)^2}   & - \frac{2 F^3 \rho_{b1} }{(1+F^2)^2 (1+BR^2)}   + \frac{2 F \rho_{b1} }{(1+F^2) (1+BR^2)}   \\
0   & 0   & -D_f k^2 - \delta_f   
\end{array}
\right) \nonumber \\
&=& 
\left(
\begin{array}{ccc}
39.8 - 0.027 k^2 & 172 & 0 \\
-97.7 & -194 - 2.7 k^2 & 9.67 \\
0   & 0   & -374 -k^2
\end{array}
\right)
\end{eqnarray}
$J$ has at least one positive eigenvalue, e.g. $k=20$,  $\lambda=15.9$. Similarly on a growing domain we find $\lambda=15.7$ for $k=20$. Therefore we can conclude that the instability is diffusion-driven and therefore of Turing type. \\

The linear stability analysis that we carried out here required multiple simplifications and the predicted wavenumber differs significantly from those observed in the numerical studies. It therefore seems that carrying out a more detailed analysis of the simplified system will not bring significant insights for the understanding of the full model.

\clearpage
\newpage

\section*{Supplementary Figures}

\renewcommand{\thefigure}{S\arabic{figure}}
\setcounter{figure}{0}

\begin{figure}[h!]
\begin{center}
\includegraphics[width=1\textwidth]{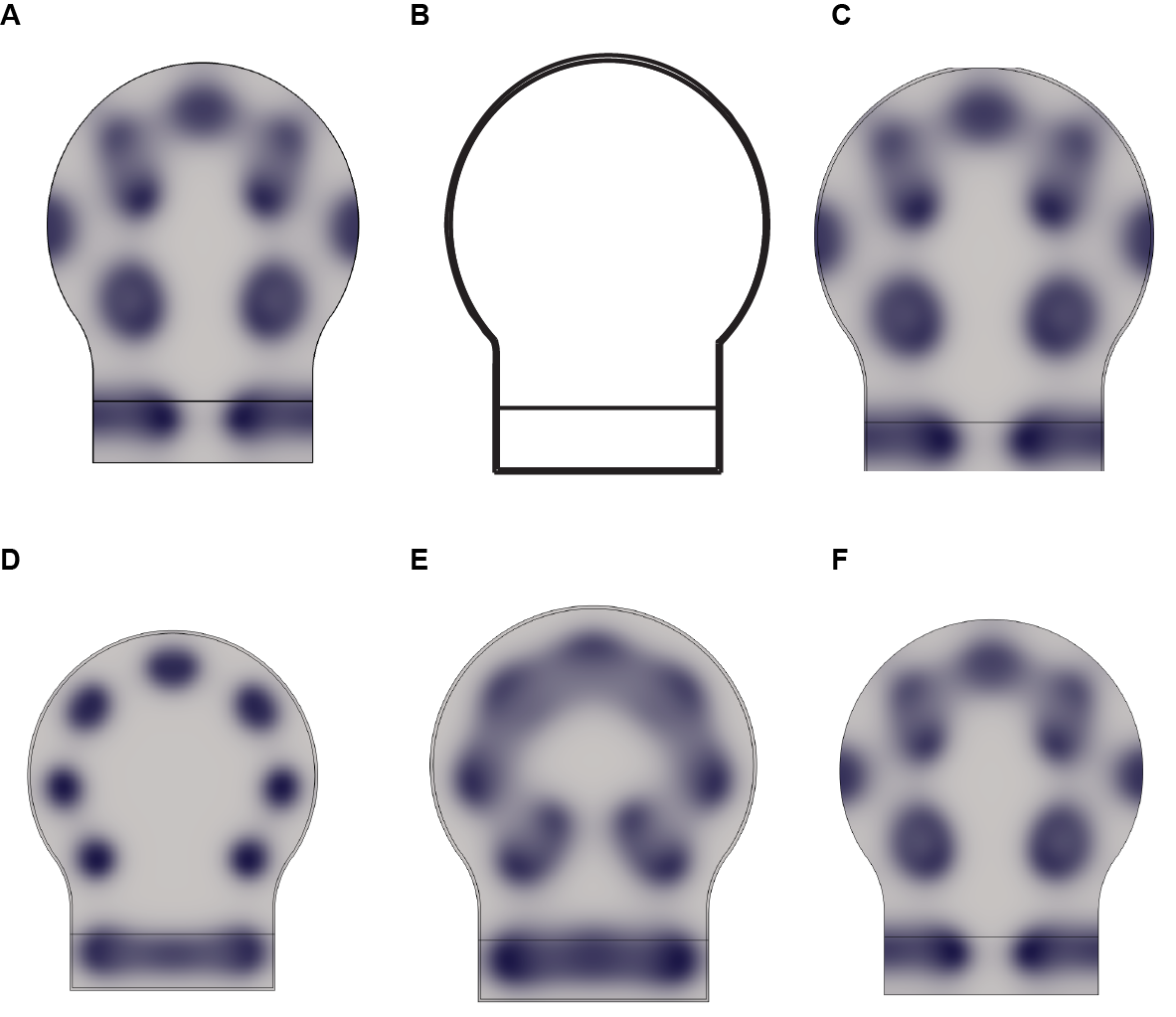}
\caption{\label{figS1} \textbf{Alternative Implementations of FGF production in the AER.} \textbf{(A)} The standard model with FGF implemented as flux. \textbf{(B-E)} A thin layer is implemented inside the boundary of the computation domain at a distance of $0.02 \cdot R_{Ref}$ as illustrated in panel B. \textbf{(C)} Thin layer with flux boundary condition.  \textbf{(D)} Thin layer where FGF production is enhanced by BMP receptor signaling. The simulation was carried out with a slightly lower ($90\%$) FGF production rate $\rho_F^{sim}=0.9 \rho_F$.  \textbf{(E)} Thin layer with constant production of FGF on a thin layer. The simulation was carried out with $\rho_F^{sim}=0.8\rho_F$ \textbf{(F)} Eq. 11 is defined on the boundary. }
\end{center}
\end{figure}

\begin{figure}[h]
\begin{center}
\includegraphics[width=1\textwidth]{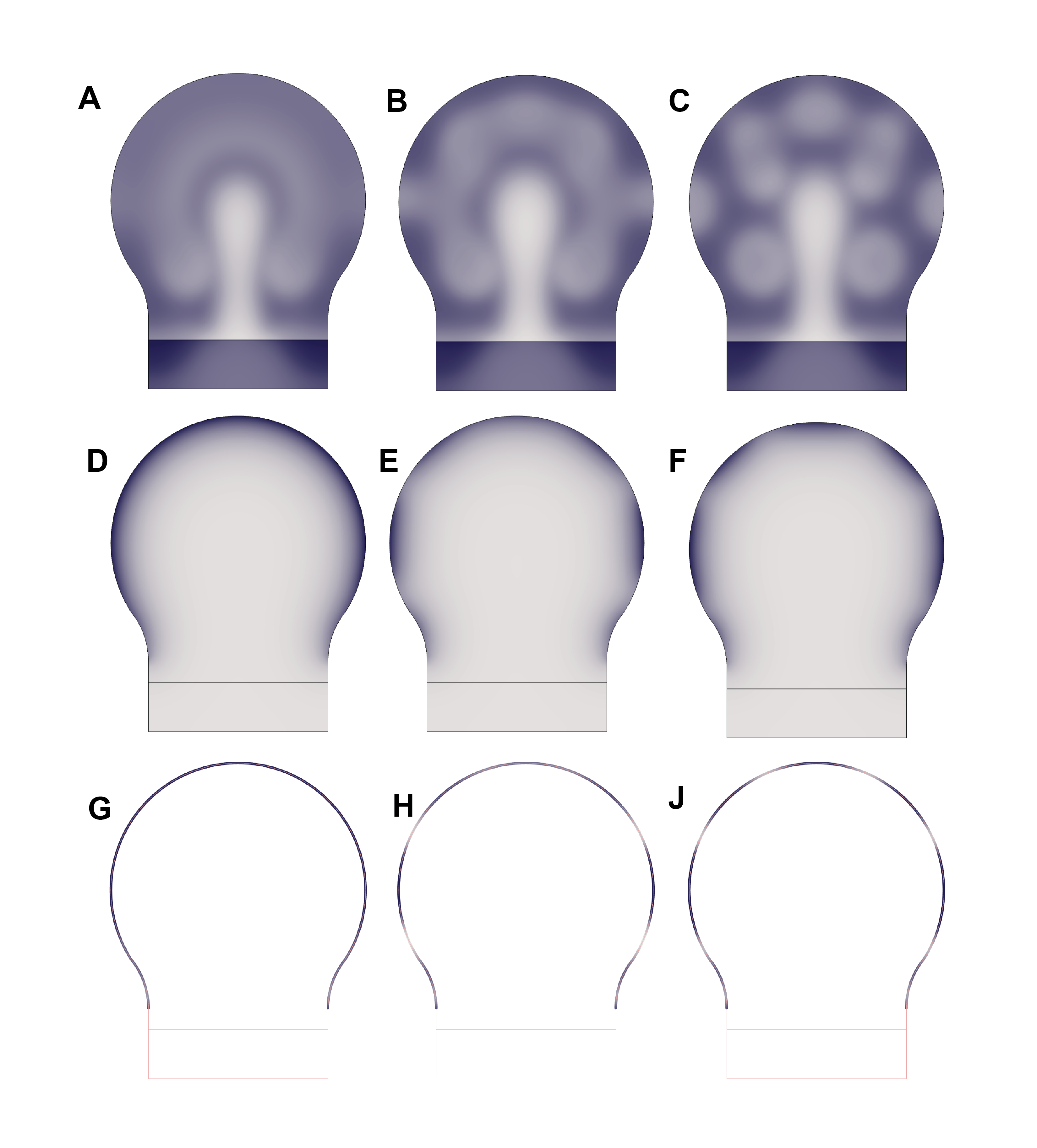}
\caption{\label{figS3} \textbf{The spatial distribution of BMP and FGF expression and activity.}   \textbf{(A-C)} The spatial distribution of BMP expression (i.e. $ \rho_B +\rho_B^* (\frac{F^n}{ F^n + 1})(\frac{1} {R^2 B + 1})  $), \textbf{(D-F)} the concentration of FGF at different simulation times,  and \textbf{(G-J)} the distribution of FGF expression on the boundary (i.e. $ \rho_F   \frac{(R^2 B)^n}{ (R^2 B)^n + \kappa_1^n} \frac{\kappa_2^n}{ (R^2 B)^n + \kappa_2^n}$) at different simulation times: \textbf{(left column)} $\tau$ = 150 , \textbf{(middle column)} $\tau$ = 350, \textbf{(right column)} $\tau$ = 750.}
\end{center}
\end{figure}

\clearpage
\begin{figure}[]
\begin{center}
\includegraphics[width=0.5\textwidth]{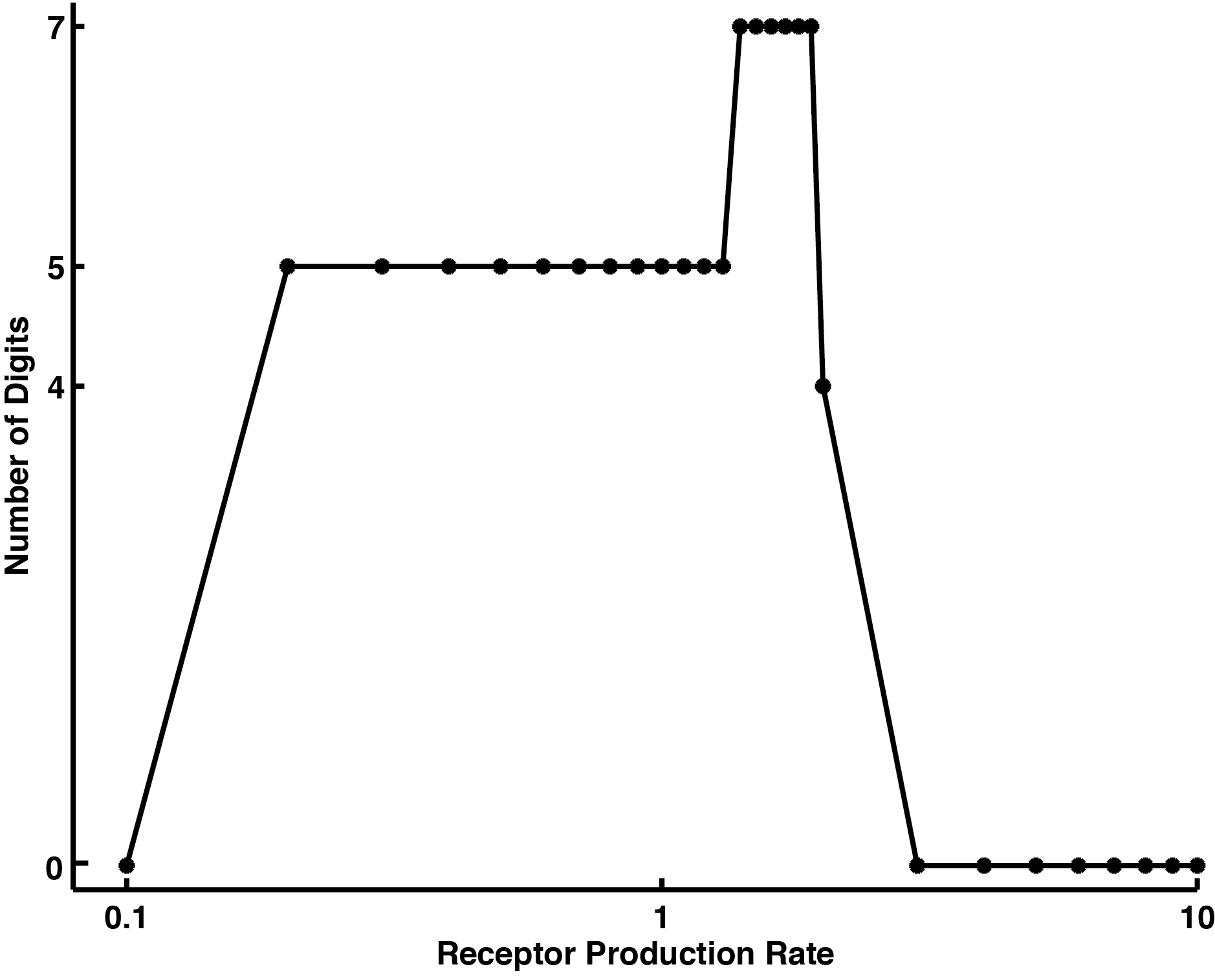}
\caption{\label{figS5} \textbf{The impact of protein production  rates on digit numbers}  Impact of the receptor production rate ($\rho_R$) on the number of spots appearing in the circular (handplate) domain. The spot(s) at the proximal end of the stalk and the two spots appearing at the transition from the stalk to the circular domain were preserved under all conditions and thus not counted. }
\end{center}
\end{figure}

\clearpage
\section*{Supplementary Tables}

\renewcommand{\thetable}{S\arabic{table}}
\setcounter{table}{0}

\begin{table}[h]
\centering
\caption{\textbf{GeometryParameters.} For an illustration see Figure 2C. $R_{Ref} $ is the reference length scale .  }\label{tbl:Geomparams}

\doublespacing
\begin{tabular}{c | c | c}
\hline
Parameter & Value & Description\\
\hline
$R_3$ & $R_{Ref}$ & Length of right semi-major axis of the elliptical limb bud  \\
$R_4$ & $R_{Ref}$ & Length of left semi-major axis of elliptical limb bud  \\
$R_1$ & $R_{Ref}$ & Length of the semi-minor axis of the elliptical limb bud   \\
$R_2$ & $-R_1*(1-W_0^2/4R_0^2)^{0.5}$ & Length between limb bud center and stalk  \\
$H_0$ & 0.77* $R_{Ref}$ & Height of the stalk   \\
$H_1$ & 0.5* $H_0$ & Height of the part of the stalk where BMP production is enhanced  ($\rho_{b2} \neq 0$) \\
$W_0$ & 1.41* $R_{Ref}$ & Width of the stalk   \\
\hline
\end{tabular}
\end{table}